\documentclass [12pt, one column, a4paper] {article}
\usepackage[utf8]{inputenc}
\usepackage{amsmath,amsfonts,amssymb}
\usepackage{bm,graphicx,graphics,color}
\usepackage{hyperref}

\newtheorem{proposition}{Proposition}[section]

\begin{document}

\title{Transition amplitude, partition function and the role of physical degrees of freedom in gauge theories}

\author{A. A. Nogueira$^{1}$\thanks{andsogueira@hotmail.com},\; B. M. Pimentel$^{2}$\thanks{pimentel@ift.unesp.br}\; and L. Rabanal$^{2}$\thanks{lrabanal@ift.unesp.br}\\
\textit{$^{1}${\small Centro de Ci\^{e}ncias Naturais e
Humanas (CCNH), Universidade Federal do ABC}}\\
\textit{\small Av. dos Estados 5001, Bairro Santa Terezinha CEP 09210-580, Santo Andr\'{e}, SP, Brazil}\\
\textit{{$^{2}${\small Instituto de F\'{i}sica Te\'orica (IFT), Universidade Estadual Paulista}}}\\
\textit{\small Rua Dr. Bento Teobaldo Ferraz 271, Bloco II Barra Funda, CEP
01140-070 S\~ao Paulo, SP, Brazil}\\
}
\maketitle
\date{}

\begin{abstract}

This work explores the quantum dynamics of the interaction between
scalar (matter) and vectorial (intermediate) particles and studies their thermodynamic equilibrium in the grand-canonical ensemble. The aim of the article is to clarify the connection between the physical degrees of freedom of a theory in both the quantization process and the description of the thermodynamic equilibrium, in which we see an intimate connection between physical degrees of freedom, Gibbs free energy and the equipartition theorem. We have split  the  work
into two sections. First, we analyze the quantum interaction in the
context of the generalized scalar Duffin-Kemmer-Petiau quantum electrodynamics (GSDKP) by using the functional formalism.  We build the Hamiltonian structure following the Dirac methodology, apply the Faddeev-Senjanovic procedure to obtain the transition amplitude in the generalized Coulomb gauge and, finally, use the Faddeev-Popov-DeWitt  method  to  write the amplitude in covariant form in the no-mixing gauge. Subsequently, we exclusively use the Matsubara-Fradkin (MF)
formalism in order to describe fields in thermodynamical equilibrium. The corresponding equations in thermodynamic equilibrium for the scalar, vectorial and ghost sectors are explicitly constructed from which the extraction of the  partition  function is straightforward. It is in the construction of the vectorial sector that the emergence and importance of the ghost fields are revealed: they eliminate the extra non-physical degrees of freedom of the vectorial sector thus maintaining the physical degrees of freedom. 
\end{abstract}

\newpage

\section{Introductory Aspects}
One of the most important parts in the analysis of physical theories is the distinction between what is measurable and what is not due to the fact that we usually use non-physical objects to describe the world \cite{Weinb}. This can be achieved by looking at the physical degrees of freedom. 

Let us review briefly the identification of the physical degrees of freedom in quantum electrodynamics in four spacetime dimensions (QED$_4$) at both zero and finite temperature where the interaction between matter (spin 1$\backslash$2, fundamental representation) and radiation (spin 1, adjoint representation) is synthesized in the following Lagrangian \cite{Ut}

\begin{equation}
\mathcal{L} = i\overline{\psi}\gamma^{\mu}D_{\mu }\psi - m\overline{\psi}\psi - \frac{1}{4}F_{\mu\nu}F^{\mu\nu},
\end{equation}
wherein the covariant derivative is defined as $D_{\mu} = \partial_{\mu} - ieA_{\mu}$ and the Dirac matrices satisfy the Clifford algebra $\gamma^{\mu}\gamma^{\nu} + \gamma^{\nu}\gamma^{\mu} = 2\eta^{\mu\nu}$.

The equivalence between the dynamics described in the phase space and the one described in the configuration space, through the Dirac constraint analysis \cite{dirac}, is contained in the following proposition

\begin{proposition}
The number of equations of motion times their order must be independent of the space in which the dynamics is described, i.e.,
\begin{equation}
	\text{(number of equations)} \times \text{(order)} = constant
\label{eq:Proposition}
\end{equation}
The conditions that must be added in order to maintain this relation are the so-called constraints.
\end{proposition}

\subsubsection*{Spinorial sector (Dirac)}

In the free spinorial sector the configuration space is described by 8 equations of first order as dictated by the Dirac equation for each component of the spinors $\psi_a$, $\overline{\psi}_a$ ($a=1,\dots,4$). However, in the phase space, we have 16 equations of first order given by the addition of the equations for the canonical momenta $p_a$, $\overline{p}_a$ ($a=1,\dots,4$), respectively. Therefore, in order to keep (\ref{eq:Proposition}) valid we must add 8 constraints
\begin{equation*}
\chi_a^{(1)},\overline{\chi}_a^{(1)} \quad\quad(a=1,\dots,4).
\end{equation*}%
On the other hand, we know that when we are dealing with physical degrees of freedom in the configuration space we have 4 wave equations of second order for the propagation of the energy of particles and anti-particles, thus, the equivalence between the dynamics in the physical degrees of freedom is maintained and no additional constraints must be added. This is, of course, a consequence of the second-class nature of the spinorial sector where gauge fixing conditions are not neccessary.

\subsubsection*{Vectorial sector (Maxwell)}

In a similar way, the vectorial sector in the configuration space is described by 4 equations of second order as dictated by the Maxwell equations for the field $A_{\mu}$
\begin{equation*}
\partial^{\mu}F_{\mu\nu} = 0, \quad\quad F_{\mu\nu} = \partial_{\mu}A_{\nu} - \partial_{\nu}A_{\mu}.
\end{equation*}
In the phase space, we have 8 equations of first order given by the incorporation of the equations for the canonical momentum $\Pi^{\mu}$. Hence, at this level, no constraints are needed. On the other hand, the physical degrees of freedom are described by 2 equations of second order due to the two polarizations (helicities) that electromagnetic waves can propagate. Therefore, in order to keep (\ref{eq:Proposition}) valid we must have 4 constraints

\begin{equation*}
\varphi _{1},\varphi _{2},\Sigma _{1},\Sigma_{2}.
\end{equation*}
The emergence of constraints in the analysis of physical degrees of freedom is, of course, a consequence of the arbitrariness in gauge theories, i.e., the first-class nature of the vectorial sector. As we can see, we have a simple and intuitive method to analyze the connection between the physical degrees of freedom and constraints in gauge theories. 

In conclusion, QED$_4$ has the following structure
\begin{equation*}
\begin{array}{c}
\text{degrees of freedom, d} \\
\\
\text{\lbrack spinorial sector (Dirac), d=4] [vectorial sector, (Maxwell), d=2]}
\\
\\
\chi ^{(1)},\overline{\chi}^{(1)},\varphi
_{1},\varphi _{2},\Sigma _{1},\Sigma _{2}
\text{ \ (12 constraints).}
\end{array}
\end{equation*}
The constraints $[\chi^{(1)},\bar{\chi}^{(1)}]$ are second-class (matter sector) and $[\varphi_{1},\varphi _{2}]$ are first-class (radiation sector). Then, $[\Sigma _{1},\Sigma _{2}]$ are the gauge fixing conditions that transform the first-class constraints into second-class constraints (Coulomb gauge) such that we can determine all the Lagrange multipliers.

This relationship between constraints and physical degrees of freedom is crucial because it is reflected in the integration measure in the quantization procedure through the functional formalism. By the Faddeev-Senjanovic method the transition amplitude, in the physical Coulomb gauge \cite{FSej}, is written in the following form
\begin{equation}
Z = N \int \mathcal{D}\mu \exp \left( i \int d^{4}x \ \left[ \left( \partial_{0}\overline{\psi} \right)p + \overline{p}\left( \partial_{0}\psi \right) + \Pi^{\nu} \left( \partial_{0}A_{\nu} \right) - \mathcal{H}_{c} \right] \right),
\end{equation}
where the integration measure is defined as
\begin{align}
\mathcal{D}\mu &= \mathcal{D}A_{\mu} \mathcal{D}\Pi^{\mu} \mathcal{D}\overline{\psi}
\mathcal{D}\psi \mathcal{D}\overline{p} \mathcal{D}p \delta(\Theta_{l}) \det\left\Vert \left\{ \Theta_{l},\Theta_{m} \right\} \right\Vert^{\frac{1}{2}}, \\
\Theta_{l} &= \left\{ \chi ^{(1)},\overline{\chi}^{(1)},\varphi _{1},\varphi _{2},\Sigma _{1},\Sigma _{2}\right\} \nonumber.
\end{align}
Next, in order to write the quantum equations of motion in an explicit covariant form we use the Faddeev-Popov-DeWitt method in the Lorenz gauge \cite{faddeev} which amounts to introduce the following identity
\begin{equation}
\det \left[ \square{\delta}^{4}(x-y) \right] \int \prod_{x} d\alpha (x) \ \delta \left[ \partial_{\mu}  {A}^{\alpha(x)\mu} \right] = 1,
\end{equation}
into the transition amplitude to write it covariantly as
\begin{align}
\label{qed1}
Z &= \tilde{N} \int \mathcal{D}A_{\mu} \mathcal{D}\overline{\psi} \mathcal{D}\psi \ \exp \left( i \int d^{4}x \ \left[ \overline{\psi} \left( i\gamma ^{\mu }D_{\mu} - m \right)\psi - \frac{1}{4}F_{\mu\nu}F^{\mu\nu} - \frac{(\partial_{\mu}A^{\mu})^2}{2\xi} \right] \right),\\
\tilde{N} &= N \det \left[ \square{\delta}^{4}(x-y) \right] \nonumber,
\end{align}
where we can write $\det \left[ \square{\delta}^{4}(x-y) \right]$ in terms of ghost fields
\begin{equation}
\det \left[ \square{\delta}^{4}(x-y) \right] = \int \mathcal{D}\overline{c} \mathcal{D}c \ \exp \left( -i \int d^{4}x \ \overline{c} \Box c \right).
\end{equation}

The choice of the Lorenz gauge is appropiated because it allows us to forget the ghosts at zero temperature due to the fact that they decouple from the vectorial sector and they can be eliminated by a suitable normalization. However, at finite temperature this statement is no longer true and the reason is simple: the global conserved charges enter into the definition of the density matrix
\begin{equation}
\hat{\rho}(\beta) = \exp \left[ -\beta \left( \hat{\mathbb{H}} - \mu_{e}\hat{N} - \mu_{g}\hat{Q} \right) \right],
\end{equation}
where $\hat{N}$ is the operator of electric charge and $\hat{Q}$ is the operator of ghost charge. In this case the partition function is given by the trace operation \cite{Mats}

\begin{align}
\label{qed2}
Z &= \text{Tr}\ \hat{\rho}(\beta) = \int \mathcal{D}A_{\mu}^E \mathcal{D}\overline{\psi} \mathcal{D}\psi \mathcal{D}\overline{c} \mathcal{D}c \mathcal{D} \ \exp \left[ -S_{T} \right], \nonumber\\
S_{T} &= \int d^{4}x\ \left[ -\frac{1}{2} A_{\mu}^E \left( \delta_{\mu\nu}\triangle + \left( 1 - \frac{1}{\xi} \right)\partial_{\mu}\partial_{\nu} \right) A_{\nu}^E - i\overline{c}\Delta c - \overline{\psi} \left( \gamma_{\mu}^{E}D_{\mu}^{(e,\mu_{e})} - m \right) \psi \right],\\
D_{\mu }^{(e,\mu_{e})} &= \partial_{\mu} - ieA_{\mu}^{E} - \mu_{e}\delta_{\mu 0}. \nonumber
\end{align}
Accordingly, in the free case ($e = 0$) we obtain
\begin{equation}
Z = Z_{\text{Maxwell}}\,Z_{\text{Dirac}}Z_{\text{Ghost}},
\end{equation}
where
\begin{align}
	Z_{\text{Maxwell}} &= \det \left[\delta_{\mu\nu}\triangle + \left( 1 - \frac{1}{\xi} \right)\partial_{\mu}\partial_{\nu} \right]^{1/2}, \\
	Z_{\text{Ghost}} &= \det \left[ \Delta \right], \label{eq:GhostDecouple}  \\
	Z_{\text{Dirac}} &= \det \left[ \gamma_{\mu}^{E}\partial_{\mu}^{(\mu_{e})} - m \right],
\end{align}
such that we can write the grand-canonical potential $\Omega$ as
\begin{eqnarray}
&&\Omega=-kT\ln Z=\Omega_{\text{Maxwell}}+\Omega_{\text{Dirac}},\cr\cr
&&\Omega_{\text{Maxwell}} = \left( \frac{1}{2}+\frac{1}{2} \right)
kT\sum_{n,\vec{p}} \ln \left[ \beta^{2} \left( {\omega_{n}}^{2}+{\vec{p}\,}^{2} \right) \right],\cr\cr
&&\Omega_{\text{Dirac}} = - \left( \frac{1}{2}+\frac{1}{2}+\frac{1}{2}+\frac{1}{2} \right)
kT\sum_{n,\vec{p}} \ln \left[ \beta^{2} \left( \left( \omega_{n}+i\mu_{e} \right)^{2}+{\vec{p}\,}^{2}+m^{2}\right) \right],
\end{eqnarray}
from which we clearly see the connection between the physical degrees of freedom and the equipartition theorem, two degrees of freedom for the photons and four for the fermions. It is worthwhile to mention that the ghost sector decouples (cf. (\ref{eq:GhostDecouple})), they eliminate the longitudinal and temporal polarizations of the vector sector maintaining the physical degrees of freedom of the theory. \cite{Abri,Vasiliev,Lands,Kapusta,Leb,Kana}

In this paper, we are interested in studying the interaction between matter (described by a Duffin-Kemmer-Petiau (DKP) field with spin 0 in the fundamental representation) and radiation (spin 1, adjoint representation \cite{Pomp}, Podolsky) described by the model known as the generalized scalar Duffin-Kemmer-Petiau quantum electrodynamics (GSDKP) \cite{AndTese}
\begin{equation}
\mathcal{L} = \frac{i}{2}\overline{\psi}\beta^{\mu}\left(\partial_{\mu}\psi\right) - \frac{i}{2}\left(\partial_{\mu}\overline{\psi}\right)\beta^{\mu}\psi - m\overline{\psi}\psi + eA_{\mu}\overline{\psi}\beta^{\mu}\psi - \frac{1}{4}F_{\mu\nu}F^{\mu\nu} + \frac{a^{2}}{2}\partial^{\mu}F_{\mu\beta}\partial_{\alpha}F^{\alpha\beta},  
\label{eq:DKP-PodolskyLagrangian}
\end{equation}
where $F_{\mu \nu }$ is the usual electromagnetic field-strength tensor, $a=\frac{1}{m_{p}}$ is the Podosky length and $\beta ^{\mu }$ are the DKP matrices that satisfy the algebra
\begin{equation}
\beta ^{\mu }\beta ^{\nu }\beta ^{\theta }+\beta ^{\theta }\beta ^{\nu
}\beta ^{\mu }=\beta ^{\mu }\eta ^{\nu \theta }+\beta ^{\theta }\eta ^{\nu
\mu }.  \label{dkpalgebra}
\end{equation}
Although DKP theory is formally
similar to Dirac theory, there are several subtleties contrasting the behaviour even at classical level \cite{Luna}. For instance, the conjugated field of the fermionic theory is characterized as $\overline{\psi} = \psi ^{\dag}\gamma ^{0}$, whereas the conjugate DKP field is defined as $\overline{\psi}=\psi ^{\dag }\eta^{0}$,
with $\eta^{0} = 2\left(\beta ^{0}\right)^{2} - 1$. Moreover, for an
arbitrary four-vector $p^{\mu}$ the following relation is satisfied
\begin{equation}
(\beta ^{\mu }p_{\mu})\left[(\beta ^{\mu }p_{\mu})^{2}-p^{2} \right] = 0,
\end{equation}
showing that $(\beta ^{\mu }p_{\mu})^{2}\neq p^{2}$, in general. Nevertheless, this relation combined with plane wave solutions of the free field equations leads to $p^{2} = m^{2}$.

The theory is also invariant under $U(1)$ global gauge transformations
\begin{equation}
\psi \rightarrow \exp \left[i\alpha \right] \psi, \qquad \overline{\psi}\rightarrow \overline{\psi}\exp \left[i\alpha \right], \qquad A_{\mu} \rightarrow A_{\mu}.
\end{equation}
Consequently, there is a conserved charge associated to it which is given by

\begin{equation}
	N = \int d^{3}x \bar{\psi}\beta ^{0}\psi, \quad\quad \dot{N} = 0.
\end{equation}
In the same form the action has also $U(1)$ local gauge symmetry
\begin{equation}
\psi \rightarrow e^{i\alpha \left( x\right) }\psi ,\quad A_{\mu }\rightarrow
A_{\mu }+\frac{1}{e}\partial _{\mu }\alpha \left( x\right) .
\end{equation}

Let us talk a bit more about the nuances of the radiation and matter sectors of GSDKP theory. When Ostrogradski constructed Lagrangian theories with higher order derivatives in classical mechanics, a new field of research was opened \cite{Ostro}. The main idea of these theories is very simple. We construct additional higher order terms such as to preserve the original symmetries of the problem, e.g., in the generalization of Utiyama work \cite{Pomp}. These theories sometimes have Hamiltonians without a lower limit \cite{PU} due to states with negative norms (ghosts), leading to the breakdown of the unitarity \cite{WH}. Attempt to restore unitarity, by avoiding ghosts, have not led to a general solution \cite{Hawk}. However, a method about how to implement terms with higher order derivatives has recently been constructed without breaking the stability of the theory. We can construct effective theories through the concept of Lagrangian anchors \cite{Russos}. It is an extension of the Noether theorems \cite{Noether} in the sense that we can also define a class of conserved quantities associated with a given symmetry.

In studying the longitudinal sector of photon propagation and its connection with mass, de Broglie suggested that the photon could be formed by a combination of two spin $\frac{1}{2}$ particles and this combination would be responsible for the photon mass \cite{broglie},

\begin{equation}
\label{combinat}
1/2\otimes 1/2=0\oplus 1.
\end{equation}
The theoretical development of this theory begins with Petiau \cite{petiau}, who gets the DKP matrix algebra (\ref{dkpalgebra}) in irreducible representations
\begin{equation}
4\otimes 4=10\oplus 5\oplus 1,
\end{equation}
in which we have 10 dimensions (representing spin 1 particles), 5 dimensions (representing spin 0 particles) and a trivial one with no physical meaning. From the previous equation we see the distribution of degrees of freedom but remembering that the Dirac sector has 4 physical degrees of freedom, the Proca sector has 3 and the scalar has 2, we will need constraints in view of proposition (\ref{eq:Proposition}). Independently, Kemmer wrote Proca equations as first order equations and did the same for Klein-Gordon-Fock (KGF) equations. From them, Kemmer conjectured about the existence of a matrix form for his system of equations \cite{Kemmer}, Duffin developed an algebra for Kemmer theory \cite{duffin} and, finally, coming from this followed the work of Kemmer \cite{kemmer}.

The DKP formalism allows us to work in a unified way the scalar and vector fields, and the wide possibility of couplings by means of covariant bilinears, unable to be described in the KGF and Proca theory, encouraged its study \cite{Umez}, in which we perceive a great phenomenological possibility in describing interactions. However, the equivalence between DKP and KGF in the free and minimally coupled cases \cite{pims}, both at the classical and quantum level, decreased the interest in DKP theory. Although the KGF formalism is apparently simpler compared to the algebraic DKP classical treatment, this point of view changes dramatically under quantization: the similarity in the form between the Lagrangians of DKP and Dirac allows us to use a very simple mechanism to study the scalar theory, since the mimic between the Dirac theory can be used to a better understanding of the physical meaning of the quantities \cite{kinoshita}. It is important to emphasize that the DKP field is usually employed in nuclear physics to describe mesons in which it is possible to say that we have a mesonic algebraic structure \cite{malg}, thus we describe bosonic fields (spin 0, spin 1) with the DKP algebra and fermionics (spin$ 1/2$) with the Clifford algebra. Yet, in describing mesons as scalar particles through electromagnetic interaction it is worth emphasizing that we are simplifying the problem. In fact, by including the electromagnetic interaction via minimum coupling together with discrete symmetries: charge conjugation, parity and time reversal (conserved by strong and electromagnetic interaction), we would have an effective Lagrangian that would describe the behavior of atomic nuclei due to the strong and electromagnetic interaction. However, the approach using DKP theory has a certain revelation when studying properties in physics, e.g., the decay of the mesons, due to their phenomenological peculiarities, and in the ratios between the strong coupling constants in the processes of interactions between two baryons with pseudo-scalar or pseudo-vectors mesons. The DKP theory matches the experimental data associated with the ratios between the strong coupling constants, whereas the theory of KGF does not \cite{decaym}. Not only in nuclear physics we see the combination of fermions (cf. (\ref{combinat})) but also in condensed matter physics, a Cooper pair is a pair of electrons (or other fermions) bound together at low temperatures in a certain manner that an arbitrarily small attraction between electrons in a metal can cause a paired state of electrons to have a lower energy than the Fermi energy, which implies that the pair is bounded. In conventional superconductors, this attraction is due to the electron–phonon interaction. Similarly, this also happens in the phenomenon of superfluidity in the description of the properties of the solitons (bosons) formed by the combination of two fermions in the liquid Helium.

Summing up, this work is devoted to the analysis of physical degrees of freedom in GSDKP. In Sec.\ref{sec1} we we establish the constraint analysis and the quantization. Although the covariant quantization is not an easy task, due to the large number of variables and the peculiar no-mixing gauge choice, after the hard work of building the transition amplitude, we can make an intuitive interpretation between the physical degrees of freedom and the constraints. In Sec.\ref{sec2} we extend the analysis to thermodynamic equilibrium following the prescription of Matsubara-Fradkin. Firstly, we extract from the equations of motion in thermodynamic equilibrium of the matter sector the partition function that would describe the interaction of DKP scalar particles with external Podolsky vectorial fields. Secondly, in the same way as the previous case, we construct the partition function of the vectorial sector that would describes the interaction between the Podolsky vectorial fields with external sources. Finally, we work in detail with the ghost sector because the ghost in the no-mixing gauge choice is described not only by grassmannian variables but also by real variables, this is singular and needed to be explored appropriately. In Sec.\ref{sec3} we unify the scalar, vectorial and ghost sectors in therms of a Gibbs variational principle of entropy. We discuss too the connection between physical degrees of freedom, Gibbs free energy and the equipartition theorem. In Sec.\ref{sec4} besides the authors present their final remarks and prospects, we see the link between the transition amplitude and the partition function, by means of Wick rotation to imaginary times.

\section{Transition amplitude and physical degrees of freedom}\label{sec1}

\subsection{Constraint analysis and the formal development}

As usual, the Euler-Lagrange equations are obtained from (\ref{eq:DKP-PodolskyLagrangian}) by the Hamiltonian
principle
\begin{eqnarray}
&&\left( i\beta ^{\mu }D_{\mu }-m \right)\psi = 0,\cr\cr
&&\left( 1+a^{2}\square \right)\partial_{\mu}F^{\lambda\mu} = e\overline{\psi}\beta
^{\lambda}\psi .
\label{eq:PodolskyEquations}
\end{eqnarray}
The translational space-time invariance of the Lagrangian density leads to the canonical Hamiltonian
\begin{align}
H_{c}=& \int d^{3}x \bigg[ \left( \partial_{0}\overline{\psi} \right)\frac{\partial\mathcal{L}}{\partial\left(\partial_{0}\overline{\psi} \right)} + \frac{\mathcal{\partial L}}{\partial (\partial _{0}\mathcal{\psi )}}(\mathcal{%
\partial }_{0}\mathcal{\psi })\mathcal{+}\frac{\mathcal{\partial L}}{%
\partial (\partial _{0}A_{\nu }\mathcal{)}}(\mathcal{\partial }_{0}A_{\nu })
\notag \\
&-\partial _{\theta }\left( \frac{\mathcal{\partial L}}{\partial (\partial
_{0}\partial _{\theta }A_{\nu }\mathcal{)}}\right) (\mathcal{\partial }%
_{0}A_{\nu }) +\frac{\mathcal{\partial L}}{\partial (\partial _{0}\partial
_{\theta }A_{\nu }\mathcal{)}}(\partial _{\theta }\mathcal{\partial }%
_{0}A_{\nu })- \mathcal{L} \bigg].
\label{CanonicalHamiltonian}
\end{align}
Thus, the canonical momenta associated with the DKP fields $\left(\overline{\psi}%
,\psi\right)$ are
\begin{align}
p &=\frac{\partial\mathcal{L}}{\partial\left(\partial_{0}\overline{\psi} \right)}%
=-\frac{i}{2}\beta ^{0}\psi, \\
\overline{p}&=\frac{\mathcal{\partial L}}{\partial (\partial _{0}\mathcal{\psi )}}%
=\frac{i}{2}\overline{\psi}\beta ^{0} ,  \label{vinpri1}
\end{align}
whereas those for gauge fields are obtained from the
Ostrogradski method \cite{Ostro}. This method consists in defining the
dynamics of the system in a first-order form, i.e., the dynamics takes place
in a spanned phase space characterized by the canonical variables $%
A_{\mu}, \Pi ^{\nu }$ and $\Gamma_{\mu} \equiv \partial_0 A_{\mu}, \Phi
^{\nu }$
\begin{align}
\Pi^{\nu} &= \frac{\partial\mathcal{L}}{\partial\Gamma_{\nu}} - 2\partial_k\frac{\partial\mathcal{L}}{\partial\left(\partial_{k}\Gamma_{\nu}\right)} - \partial_{0}\frac{\partial\mathcal{L}}{\partial\left(\partial_{0}\Gamma_{\nu}\right)} = F^{\nu0} + a^{2}\left[\eta^{i\nu}\partial_{i}\partial_{\alpha}F^{\alpha 0} - \partial_{0}\partial_{\alpha}F^{\alpha\nu}\right], \\
\Phi^{\nu} &= \frac{\partial\mathcal{L}}{\partial\left(\partial_{0}\Gamma_{\nu}\right)} = a^{2}\left[\partial_{\alpha}F^{\alpha\nu} - \eta^{\nu 0}\partial_{\alpha}F^{\alpha 0}\right].  \label{vinpri2}
\end{align}
From the above momentum expressions, the constraint structure
of the theory can be studied by the Dirac's treatment of singular systems \cite{dirac}. In this way, the set of first-class constraints
\begin{align}
\varphi _{1}=\Phi _{0}\approx 0 , \quad \varphi _{2}= \Pi _{0}-\partial
_{k}\Phi ^{k}\approx 0 , \quad \varphi _{3}= e\overline{\psi}\beta
^{0}\psi-\partial ^{k}\Pi _{k}\approx 0 ,
\end{align}
and the set of second-class constraints
\begin{align}
\chi ^{(1)}&=p+\frac{i}{2}\beta ^{0}\psi \approx 0 , \quad \overline{\chi}^{(1)} =%
\overline{p}-\frac{i}{2}\overline{\psi}\beta ^{0}\approx 0 , \\
\chi ^{(2)}& = \left[1-\left(\beta ^{0}\right)^{2}\right]\left[i\beta ^{i}\partial _{i}\psi (x)-m\psi
(x)+e\beta ^{i}A_{i}(x)\psi (x)\right]\approx 0 , \\
\overline{\chi}^{(2)} &= \left[-i\partial _{i}\overline{\psi}(x)\beta ^{i}+m\overline{\psi}(x)-e%
\overline{\psi}(x)\beta ^{i}A_{i}(x)\right]\left[1-\left(\beta ^{0}\right)^{2}\right]\approx 0,
\end{align}
are obtained. The weak equality $\approx $ is understood in according to Dirac's sense.

Having determined the full set of constraints, the next step is to obtain the functional generator. The transition
amplitude in the Hamiltonian form is written in the following way \cite{FSej}
\begin{equation}
Z = N\int \mathcal{D}\mu \ \exp \left( i\int d^{4}x \ \left[ \left( \mathcal{\partial }_{0}
\mathcal{\overline{\psi}}\right) p+\overline{p}\left( \mathcal{\partial }_{0}\mathcal{%
\psi }\right) \mathcal{+}\Pi ^{\nu }\left( \mathcal{\partial }_{0}A_{\nu
}\right) +\Phi ^{\nu }\left( \partial _{0}\Gamma _{\nu }\right) \mathcal{-H}%
_{c}\right] \right),
\end{equation}%
where the canonical Hamiltonian density is given by
\begin{align}
\mathcal{H}_{c} &= \Pi _{0}\Gamma ^{0}+\Pi _{k}\Gamma ^{k} + \Phi _{k} \left( \partial
^{k}\Gamma _{0}-\partial _{l}F^{lk}+\frac{\Phi ^{k}}{2a^{2}} \right) - \frac{i}{2}%
\overline{\psi}\beta ^{i}\overleftrightarrow{\partial_i}\psi +m\overline{\psi}\psi
\nonumber \\
&\quad-e\overline{\psi}\hat{A}\psi +\frac{1}{4}F_{kj}F^{kj}+\frac{1}{4}\left(\Gamma
_{j}-\partial _{j}A_{0}\right)^{2}-\frac{a^{2}}{2}\left(\partial ^{j}\Gamma
_{j}-\partial ^{j}\partial _{j}A_{0}\right)^{2},
\end{align}
and the integration measure is defined such that it transforms as
a scalar in the constrained phase space
\begin{equation}
\mathcal{D}\mu = \mathcal{D}\Phi ^{\nu } \mathcal{D}\Gamma _{\nu } \mathcal{D}\Pi ^{\mu } \mathcal{D}A_{\mu } \mathcal{D}\mathcal{\overline{\psi}}%
\mathcal{D}\mathcal{\psi } \mathcal{D}\overline{p} \mathcal{D}p\delta (\Theta _{l})\det \left\Vert \left\{\Theta
_{l},\Theta _{m} \right\}\right\Vert ^{1/2}.
\end{equation}%
We will derive the complete set of constraints for the GSDKP model in the next subsection. It is expressed as
\begin{equation}
\Theta _{l} = \left\{ \chi ^{(1)},\overline{\chi}^{(1)},\chi ^{(2)},\overline{\chi}%
^{(2)},\varphi _{1},\varphi _{2},\varphi _{3},\Sigma _{1},\Sigma _{2},\Sigma
_{3}\right\} ,
\end{equation}
in which suitable gauge fixing conditions are chosen \cite{galvao}
\begin{equation}
\Sigma _{1}=\Gamma _{0}(x)\approx 0, \quad \Sigma _{2}=A_{0}\approx 0, \quad
\Sigma _{3} = \left(1+a^{2}\square\right)\left(\vec{\nabla}\cdot\vec{A}\right)\approx 0.
\end{equation}
After integrating over the gauge and fermionic momenta, the transition
amplitude becomes
\begin{align}
Z& = N\int \mathcal{D}A_{\mu} \mathcal{D}\mathcal{\overline{\psi}} \mathcal{D}\mathcal{\psi } \ \det \left\Vert
\left(1+a^{2}\vec{\nabla}^{2}\right)\vec{\nabla}^{2}\right\Vert \delta \left(\left(1+a^{2}\square
\right)\left(\vec{\nabla}\cdot\vec{A}\right)\right)  \notag \\
&\quad\times \exp \left[ i\int d^{4}x \ \left(\mathcal{\overline{\psi}}\left(i\beta ^{\mu }\nabla _{\mu
}-m\right)\mathcal{\psi }-\frac{1}{4}F_{\mu \nu }F^{\mu \nu }+\frac{a^{2}}{2}%
\partial ^{\mu }F_{\mu \beta }\partial _{\alpha }F^{\alpha \beta }\right) \right].
\label{amplicg}
\end{align}
Although the above expression is correct its form is not explicitly
covariant, thus, it is not convenient for computations. However, the Faddeev-Popov-DeWitt ansatz \cite{faddeev}
\begin{equation}
\det \left[\left(1+a^{2}\square \right)^{1/2}\square {\delta }^{4}(x-y)\right] \int
\prod_{x}d\alpha (x) \ \delta \left[ \left(1+a^{2}\square \right)^{1/2}{%
\partial _{\mu }}{A}^{\alpha\mu}\right] = 1,
\end{equation}
allows us to transform the vacuum-vacuum transition amplitude into a covariant form. Hence, using it in the no-mixing gauge condition the transition amplitude can be written as
\begin{align}
Z &= \tilde{N} \det\left[\left(1 + a^{2}\square\right)^{1/2}\square\delta^{4}(x-y)\right] \int \mathcal{D}A_{\mu} \mathcal{D}\mathcal{\overline{\psi}} \mathcal{D}\mathcal{\psi} \ \exp \bigg[ i\int d^{4}x \ \bigg( \mathcal{\overline{\psi}}\left(i\beta^{\mu}\nabla_{\mu} - m\right)\mathcal{\psi} - \frac{1}{4}F_{\mu \nu }F^{\mu \nu }  \notag \\
&\quad+ \frac{a^{2}}{2}\partial ^{\mu }F_{\mu \beta }\partial _{\alpha }F^{\alpha
\beta } - \frac{1}{2\xi}\left( \partial ^{\mu }A_{\mu }\right) \left(
1+a^{2}\square \right) \left( \partial ^{\mu }A_{\mu }\right) \bigg) \bigg],
\end{align}
where we can write $\det\left[\left(1 + a^{2}\square\right)^{1/2}\square\delta^{4}(x-y)\right]$ in terms of ghost fields
\begin{equation}
\det\left[\left(1 + a^{2}\square\right)^{1/2}\square\delta^{4}(x-y)\right] = \int \mathcal{D}\overline{c} \mathcal{D}c \mathcal{D}\phi \ \exp \left( -i[\int d^{4}x \ \overline{c}\left(1+a^{2}\square\right)\square c+\phi\left(1+a^{2}\square\right)\phi]\right),
\end{equation}
with $\bar{c}$, $c$ grasmanian fields and $\phi$ a real field. Although, initially, Podolsky had used the Lorenz condition $(\partial _{\mu }A^{\mu })$ to fix the physical degrees of freedom, after a rigorous study involving constraint analysis it was shown that it was not the more appropiate one \cite{galvao}. As a result, the natural way of fixing the degrees of freedom has become the generalized Lorenz condition $\left( 1 + a^{2}\square\right)\left(\partial_{\mu }A^{\mu }\right)$. However, there is a price to pay, namely, this condition increases the order of the derivatives in the Lagrangian. On the other hand, there is another gauge condition known as the no-mixing gauge $\left( 1+a^{2}\square\right)^{1/2}\partial ^{\mu }A_{\mu }$ that combines perfectly with the Podolsky theory maintaining the order of the Lagrangian \cite{Daniel,pdif,LW}. But, in this case it is necessary to deal with a pseudo-differential structure. Despite of the peculiarities of each gauge choice, the Lorenz condition, no-mixing and generalized Lorenz are related by BRST symmetry \cite{And}.

The minimal coupling DKP functional generator with the higher-derivative
Podolsky term can be written as
\begin{equation}
\mathcal{Z}\left[ \eta ,\bar{\eta},J_{\mu }\right] = \int \mathcal{D}\mu \left(\psi,\overline{\psi},A_{\mu},\overline{c},c,\phi\right) \exp \left[ iS_{\text{eff}}\right],  \label{funcgerador}
\end{equation}
with the effective action given by
\begin{align}
S_{\text{eff}} &= \int d^{4}x \ \bigg[ \overline{\psi} \left(i\beta^{\mu}\partial_{\mu} - m + e\beta^{\mu}A_{\mu}\right)\psi - \frac{1}{4}F_{\mu\nu}F^{\mu\nu} + \frac{a^{2}}{2}\partial^{\mu}F_{\mu\beta}\partial_{\alpha}F^{\alpha\beta} \notag \\
&\quad+ \frac{1}{2\xi}\left(\partial^{\mu}A_{\mu}\right)\left(1 + a^{2}\square\right)\left(\partial^{\mu}A_{\mu}\right) - \overline{c}\left(1 + a^{2}\square\right)\square c - \phi\left(1 + a^{2}\square\right)\phi \notag \\
&\quad+ \overline{\psi}\eta + \overline{\eta}\psi + A^{\mu}J_{\mu} + \overline{c}\zeta + \overline{\zeta}c + \phi J \bigg],
\label{acao}
\end{align}
where $\eta$, $\overline{\eta}$, $J_{\mu }$, $\overline{\zeta}$, $\zeta$ and $J$ are the sources for the fields $\psi$, $\overline{\psi}$, $A_{\mu }$, $c$, $\overline{c}$ and $\phi$, respectively. From here it follows that the Schwinger-Dyson
equations and Ward-Takahashi identities can straightforwardly be obtained \cite{Proceedings}.

\subsection{The physical degrees of freedom and the connection with constraints}\label{sec:GSDKP-Constraints}

This section is devoted to the identification of the physical degrees of freedom and to the determination of the constraints according to proposition (\ref{eq:Proposition}). The analysis mimics the one we did in the introduction.

\subsubsection*{Scalar sector (DKP)}

In the scalar sector the configuration space is described by 10 equations of first order as dictated by the DKP equations for each component $\psi_a$, $\overline{\psi}_a$ ($a=1,\dots,5$). However, in the phase space, we have 20 equations of first order given by the addition of the equations for the canonical momenta $p_a$, $\overline{p}_a$ ($a=1,\dots,5$), respectively. Therefore, in order to keep (\ref{eq:Proposition}) valid we must add 10 constraints
\begin{equation*}
\chi_a^{(1)},\overline{\chi}_a^{(1)} \quad\quad(a=1,\dots,5).
\end{equation*}%
On the other hand, we know that when we are dealing with physical degrees of freedom in the KGF equation we have 2 equations of second order, namely, the equations for $\phi$ and its complex conjugate $\phi^*$. Therefore, in order to keep (\ref{eq:Proposition}) valid we need to have 16 constraints in total. Thus, it amounts to add 6 additional constraints
\begin{equation*}
\chi ^{(2)}_b,\overline{\chi}^{(2)}_b \quad\quad (b = 1,2,3).
\end{equation*}

\subsubsection*{Vectorial sector (Podolsky)}
In a similar way, the vectorial sector in the configuration space is described by 4 equations of fourth order as dictated by the Podolsky equations (\ref{eq:PodolskyEquations}). In the phase space, we have 16 equations of first order given by the canonical pairs $A_{\mu }$, $\Pi ^{\mu }$ and $\Gamma ^{\nu }$, $\Phi _{\nu }$. Hence, at this level, no constraints are needed. On the other hand, the physical degrees of freedom are described by 5 equations of second order due to the two and three polarizations of Maxwell and Proca particles, respectively. Therefore, in order to keep (\ref{eq:Proposition}) valid we must have 6 constraints

\begin{equation*}
\varphi _{1},\varphi _{2},\varphi_{3},\Sigma _{1},\Sigma
_{2},\Sigma _{3}.
\end{equation*}

In conclusion, GSDKP has the following structure
\begin{equation*}
\begin{array}{c}
\text{degrees of freedom, d} \\
\\
\text{\lbrack scalar sector (DKP), d=2] [vectorial sector (Podolsky) d=2+3]}
\\
\\
\chi ^{(1)},\overline{\chi}^{(1)},\chi ^{(2)},\overline{\chi}^{(2)},\varphi
_{1},\varphi _{2},\varphi_{3},\Sigma _{1},\Sigma _{2},\Sigma _{3}
\text{ \ (22 constraints).}
\end{array}
\end{equation*}

Again, this relationship between constraints and physical degrees of freedom is crucial, and it is reflected both in its quantization of the GSDKP, as we have seen in the previous section, and in describing its thermodynamic equilibrium, by means of the free energy and the equipartition theorem, as we shall see now.

\section{GSDKP in thermal equilibrium}\label{sec2}

To construct the partition function for GSDKP it will be necessary to study separately the matter, radiation and ghost sectors to later unify them in a variational principle of maximum Gibbs entropy.

\subsection{The matter sector}
Firstly, we will study the equations of motion in the Hamiltonian formalism. From previous discussions the canonical Hamiltonian of the scalar sector in the presence of an external field is given by

\begin{equation}
H_{c} = \int d^{3}x \ \left[ -\frac{i}{2}\overline{\psi}\beta^{i} \left(
\overleftrightarrow{\partial_i}\psi \right) + m\overline{\psi}\psi - eA_{\mu}\overline{\psi}\beta^{\mu}\psi \right].
\end{equation}
Consequenlty, the equation of motion for $\beta^{0}\psi$ is given by\footnote{The Dirac brackets are defined as
\begin{equation}
	\left\{ F, G\right\}_D = \left\{ F, G \right\}_P + \left\{ F,\Psi^{\alpha} \right\}_P \left\{ \Psi_{\alpha}, \Psi_{\beta} \right\}^{-1}_P \left\{ \Psi_{\beta}, G \right\}_P
\end{equation}
for any function $F$ and $G$ in the phase space and $\Psi=\{\chi ^{(1)},\overline{\chi}^{(1)},\chi ^{(2)},\overline{\chi}^{(2)}\}$ representing the set of second-class constraints.}

\begin{align}
\frac{d}{dt} \left( \beta^{0}\psi \right) &= \int d^{3}z \ \left\{ \beta^{0}\psi, \mathcal{H}_{c}(z) \right\}_{D} \nonumber \\
&= \int d^{3}z \ \left\{ \beta ^{0}\psi, -\frac{i}{2} \overline{\psi}(z)\beta^{i} \partial_{i}\psi(z) + \frac{i}{2} \partial_{i}\overline{\psi}(z)\beta^{i}\psi(z) + m\overline{\psi}(z)\psi(z) - eA_{\mu}(z)\overline{\psi}(z)\beta^{\mu }\psi(z) \right\}_{D}.
\label{eqdin}
\end{align}
The equivalence between the Hamiltonian dynamics, the Lagrangian dynamics and the analysis of the second-class constraints lead us to the result
\begin{equation}
\left\{\beta ^{0}\psi (x),\overline{\psi}(y)\right\}_{D} = -iI\delta ^{3}(\vec{x}-\vec{y}).
\end{equation}%
Therefore, (\ref{eqdin}) yields the following equation of motion
\begin{equation}
\left[ i\beta^{\mu}\left(\partial _{\mu }-ieA_{\mu }\right) - m \right]\psi =0.
\end{equation}
Analogously, we find the equation of motion for $\overline{\psi}\beta^{0}$ through
\begin{align}
\frac{d}{dt} \left( \overline{\psi}\beta ^{0} \right) &= \int d^{3}z \ \left\{ \overline{\psi}\beta^{0},\mathcal{H}_{c}(z)\right\}_{D} \nonumber\\
&= \int d^{3}z \ \left\{ \overline{\psi}\beta ^{0}, -\frac{i}{2}\overline{\psi}(z)\beta^{i}\partial _{i}\psi(z) + \frac{i}{2}\partial _{i}\overline{\psi}%
(z)\beta ^{i}\psi (z) + m\overline{\psi}(z)\psi (z)-eA_{\mu }(z)\overline{\psi}(z)\beta ^{\mu }\psi (z)\right\}_{D},%
\end{align}%
using
\begin{equation}
\left\{\overline{\psi}\beta ^{0},\psi (y) \right\}_{D} = iI\delta ^{3}(\vec{x}-\vec{y}),
\end{equation}%
we obtain
\begin{equation}
\overline{\psi} \left[ i\beta^{\mu} \left(\overleftarrow{\partial_{\mu}} + ieA_{\mu}\right) + m\right] = 0.
\end{equation}

Using the Dirac correspondence principle, $\{\cdot,\cdot\}_{D}\rightarrow -i[\cdot,\cdot]$, we quantize the theory
\begin{equation}
\left[ \beta^{0}\hat{\psi}(x),\hat{\overline{\psi}}(y) \right] = I\delta ^{3}(\vec{x}-\vec{y}) \qquad \left[ \hat{\overline{\psi}}\beta^{0}, \hat{\psi}(y) \right] = -I\delta^{3}(\vec{x} - \vec{y}).
\label{quantc}
\end{equation}%
Because we are concerned with the study of thermodynamic equilibrium, the above equations become stationary in the sense that they do not depend on time. At this point, we can define the density matrix in the grand canonical ensemble and describe the scalar sector of GSDKP electrodynamics with sources. The density matrix has the form

\begin{equation}
	\hat{\rho}_{s}(\beta) = \exp \left[ -\beta\left(\hat{H} - \mu_{e}\hat{N}\right)\right],
\label{m}
\end{equation}
with

\begin{align}
	\hat{H} &= \int d^{3}z \ \left( -\frac{i}{2}\hat{\overline{\psi}}\beta^{i}\partial_{i}\hat{\psi} + \frac{i}{2}\partial_{i}\hat{\overline{\psi}}\beta^{i}\hat{\psi} + m\hat{\overline{\psi}}\hat{\psi} - eA_{\mu}\hat{\overline{\psi}}\beta^{\mu}\hat{\psi} + \overline{\eta}\hat{\psi} + \hat{\overline{\psi}}\eta \right), \nonumber\\
	\hat{N} &= \int d^{3}z \ \hat{\overline{\psi}}\beta^{0}\hat{\psi} \nonumber.
\end{align}
Applying the similarity transformation $\hat{\mathcal{O}}^{s} = \hat{\rho}_{s}^{-1}\hat{\mathcal{O}}\hat{\rho}_{s}$ in (\ref{quantc}) we obtain the commutators in thermodynamic equilibrium
\begin{eqnarray}
&&\left[ \beta ^{0}\hat{\psi^{s}} \left(\vec{x},\tau\right),\hat{\overline{\psi^s}}\left(\vec{y},\tau\right)\right] = I\delta ^{3}(\vec{x}-\vec{y}),\cr\cr
&&\left[ \hat{\overline{\psi^s}} \left(\vec{x},\tau \right)\beta^{0},\hat{\psi^{s}} \left(\vec{y},\tau \right)\right] = -I\delta ^{3}(\vec{x}-\vec{y}).
\end{eqnarray}

Next, we determine the equations of motion in thermodynamic equilibrium. For $\beta^{0}\hat{\psi^{s}}$ we have
\begin{align}
\frac{\partial}{\partial \tau} \left(\beta^{0}\hat{\psi^{s}}\right) &=%
\frac{\partial}{\partial \tau } \bigg( e^{\tau \left(\hat{H} - \mu_{e}\hat{N}\right)}\beta^{0}\hat{\psi} e^{-\tau \left(\hat{H} - \mu_{e}\hat{N}\right)} \bigg), \\
&=e^{\tau \left(\hat{H} - \mu_{e}\hat{N}\right)} \left[\hat{H} - \mu_{e}\hat{N}, \beta^{0}\hat{\psi}\right]e^{-\tau \left(\hat{H} - \mu_{e}\hat{N}\right)},  \nonumber
\end{align}
or
\begin{align}
\frac{\partial}{\partial \tau } \left( \beta^{0}\hat{\psi^{s}} \right)%
&=\int d^{3}z \ \bigg[ -\frac{i}{2}\hat{\psi^{s}}\beta^{i}\partial_{i}\hat{\psi^{s}} + \frac{i}{2}\partial _{i}\hat{\overline{\psi^s}}\beta^{i}\hat{\psi^{s}} + m\hat{\overline{\psi^s}}\hat{\psi^{s}} - eA_{\mu}\hat{\overline{\psi^s}}\beta^{\mu}\hat{\psi^{s}} - \mu_{e}\hat{\overline{\psi^s}}\beta^{0}\hat{\psi^{s}} \nonumber\\ 
&\quad+\overline{\eta}\hat{\psi^{s}} + \hat{\overline{\psi^s}}\eta, \beta^{0}\hat{\psi} \bigg], \\
\eta &= -\beta^{0}\partial_{0}\hat{\psi^{s}} + i\beta^{j}\partial_{j}\hat{\psi^{s}} + eA_{\mu}\beta^{\mu }\hat{\psi^{s}} + \mu_{e}\beta^{0}\hat{\psi^{s}} - m\hat{\psi^{s}}.%
\end{align}
The Euclidean structure of the algebra appears after the definitions
\begin{equation}
\beta _{0}^{E} = -\beta^{0}, \qquad \beta_{j}^{E} = i\beta^{j}, \qquad \beta_{\mu}^{E}\beta_{\nu}^{E}\beta_{\theta}^{E} + \beta_{\theta}^{E}\beta_{\nu}^{E}\beta_{\mu}^{E} = \beta_{\mu}^{E}\delta_{\nu\theta} + \beta_{\theta}^{E}\delta_{\nu\mu}.
\end{equation}%
Moreover, in order to write the covariant equation of motion in thermodynamic equilibrium we define
\begin{equation}
A_{0}^{E} = -iA_{0}, \qquad A_{j}^{E} = A_{j},
\end{equation}
such that
\begin{equation*}
A_{\mu}\beta^{\mu} = A_{0}\beta^{0} + A_{j}\beta^{j} = -i\left( A_{0}^E\beta_{0}^{E} + A_{j}^E\beta_{j}^{E} \right) = -i A_{\mu}^E\beta_{\mu}^E.
\end{equation*}
All in all, we obtain the following equation

\begin{equation*}
\left[ \beta_{\mu}^{E} D_{\mu}^{\left(e,\mu_{e}\right)} -m\right]\hat{\psi^{s}} = \eta,
\end{equation*}
with $D_{\mu }^{\left(e,\mu_{e}\right)} = \partial _{\mu }-ieA_{\mu}^{E}-\mu _{e}\delta _{\mu 0}$. In a similar way, we can compute the equation of motion for $\hat{\overline{\psi^s}}\beta ^{0}$ to, finally, end up with the thermal equations for the scalar sector with external sources
\begin{align}
\left[ \beta_{\mu}^{E} D_{\mu}^{\left(e,\mu_{e}\right)} -m\right]\hat{\psi^{s}} &= \eta, \label{eq:Scalar1}\\
\hat{\overline{\psi^s}}\left[\overleftarrow{D}_{\mu}^{\left(-e,-\mu_{e}\right)}\beta_{\mu}^{E} + m\right] &= \overline{\eta}.%
\label{seq}
\end{align}

In order to find a functional integral representation for these equations note that they can be rewritten with the help of the density matrix (\ref{m}) as

\begin{align}
\left[ \left(\beta_{\mu}^{E}\right)_{ab} \overrightarrow{\partial_{\mu }^{\mu _{e}}} - m\delta_{ab} \right] \frac{\delta\hat{\rho_{s}}(\beta)}{\delta\overline{\eta}_{b}\left(\vec{x},\tau\right)} &= ie\left(\beta_{\mu}^{E}\right)_{ab} \frac{\delta\hat{\rho_{s}}(\beta)}{\delta\overline{\eta}_{b}\left(\vec{x},\tau\right)}A_{\mu}^{E}+ \eta\hat{\rho_{s}}(\beta), \nonumber \\
\frac{\delta\hat{\rho_{s}}(\beta)}{\delta\eta_{a}\left(\vec{x},\tau\right)} \left[\left(\beta_{\mu}^{E}\right)_{ab}\overleftarrow{\partial_{\mu}^{-\mu_{e}}} + m\delta_{ab}\right] &= -ie\left(\beta_{\mu}^{E}\right)_{ab}\frac{\delta\hat{\rho_{s}}(\beta)}{\delta\eta_{b}\left(\vec{x},\tau\right)}A_{\mu}^{E} + \hat{\rho_{s}}(\beta)\overline{\eta}.
\end{align}
After taking the trace of the previous expressions with the purpose of finding equations for average values we establish the functional equations satisfied by the generating functional $Z_{1}\left[\mathcal{J},\overline{\eta},\eta\right]$
\begin{align}
\left[ \left(\beta_{\mu}^{E}\right)_{ab} \overrightarrow{\partial_{\mu }^{\mu _{e}}} - m\delta_{ab} \right] \frac{\delta Z_{1}\left[\mathcal{J},\overline{\eta},\eta\right]}{\delta\overline{\eta}_{b}\left(\vec{x},\tau\right)} &= ie\left(\beta_{\mu}^{E}\right)_{ab} \frac{\delta Z_{1}\left[\mathcal{J},\overline{\eta},\eta\right]}{\delta\overline{\eta}_{b}\left(\vec{x},\tau\right)}A_{\mu}^{E}+ \eta Z_{1}\left[\mathcal{J},\overline{\eta},\eta\right], \nonumber \\
\frac{\delta Z_{1}\left[\mathcal{J},\overline{\eta},\eta\right]}{\delta\eta_{a}\left(\vec{x},\tau\right)} \left[\left(\beta_{\mu}^{E}\right)_{ab}\overleftarrow{\partial_{\mu}^{-\mu_{e}}} + m\delta_{ab}\right] &= -ie\left(\beta_{\mu}^{E}\right)_{ab}\frac{\delta Z_{1}\left[\mathcal{J},\overline{\eta},\eta\right]}{\delta\eta_{b}\left(\vec{x},\tau\right)}A_{\mu}^{E} + Z_{1}\left[\mathcal{J},\overline{\eta},\eta\right](\beta)\overline{\eta}.
\label{sfunc}
\end{align}
Let us introduce the ansatz for the solution of the equations (\ref{sfunc}) by standard Fourier transformations techniques as
\begin{equation}
Z_{1}[\bar{\eta},\eta ] = \int \mathcal{D}\overline{\psi} \mathcal{D}\psi \ \widetilde{Z}_{1}\left[\mathcal{A},\overline{\psi},\psi\right]\exp \left(\overline{\eta}\psi +\overline{\psi}\eta \right).%
\end{equation}%
Thus, from (\ref{sfunc}) we obtain

\begin{align}
\int \mathcal{D}\overline{\psi} \mathcal{D}\psi \left[ \beta_{\mu}^{E} D_{\mu}^{\left(e,\mu_{e}\right)} -m\right]\psi \widetilde{Z}_{1}\left[\mathcal{A},\overline{\psi},\psi\right]\exp \left(\overline{\eta}\psi +\overline{\psi}\eta \right) &= \eta Z_{1}\left[\mathcal{J},\overline{\eta},\eta \right], \nonumber\\
\int \mathcal{D}\overline{\psi} \mathcal{D}\psi \ \widetilde{Z}_{1}\left[\mathcal{A},\overline{\psi},\psi\right]\exp \left(\overline{\eta}\psi +\overline{\psi}\eta \right) \left[\overleftarrow{D}_{\mu}^{\left(-e,-\mu_{e}\right)}\beta_{\mu}^{E} + m\right] &= Z_{1}\left[\mathcal{J},\overline{\eta},\eta \right]\overline{\eta}.%
\end{align}%
Defining the quantity
\begin{equation}
S_{1} = -\int d^{4}x \ \overline{\psi}\left(\vec{x},\tau\right)\left[ \beta_{\mu}^{E} D_{\mu}^{\left(e,\mu_{e}\right)} -m\right]\psi\left(\vec{x},\tau\right),
\end{equation}
such that
\begin{align}
\frac{\delta S_{1}}{\delta \overline{\psi}\left(\vec{x},\tau\right)} &= -\left[ \beta_{\mu}^{E} D_{\mu}^{\left(e,\mu_{e}\right)} -m\right]\psi\left(\vec{x},\tau\right), \nonumber \\
\frac{\delta S_{1}}{\delta \psi\left(\vec{x},\tau\right)} &= \overline{\psi}\left(\vec{x},\tau\right)\left[\beta _{\mu }^{E}\overleftarrow{D}_{\mu }^{(-e,-\mu _{e})}+m\right],%
\end{align}%
we realize that the partition function of the scalar sector in the presence of an external field is 
\begin{equation}
Z_{1}\left[\mathcal{J},\overline{\eta},\eta \right] = \int \mathcal{D}\overline{\psi} \mathcal{D}\psi \ \exp \left(-S_{1}+\overline{\eta}\psi+\overline{\psi}\eta \right).
\end{equation}

\subsection{The vectorial sector}

We begin this section with the Lagrangian density that describes the free vectorial sector with the covariant no-mixing gauge fixing term
\begin{equation}
\mathcal{L} = -\frac{1}{4}F_{\mu \nu }F^{\mu \nu } + \frac{1}{2m_{p}^{2}}\partial _{\mu }F^{\mu \lambda }\partial ^{\theta }F_{\lambda
\theta } - \frac{1}{2\xi }\partial _{\mu }A^{\mu }\left(1+\frac{\square }{m_{p}^{2}}\right)\partial _{\nu }A^{\nu }.
\end{equation}
Using (\ref{CanonicalHamiltonian}) we determine the canonical Hamiltonian. To do that, we introduce the canonically conjugate momenta within the Ostrogradski method (cf. (\ref{vinpri2}))
\begin{align}
\Pi^{\nu} &= F^{\nu 0} + \frac{1}{m_{p}^{2}} \left(\eta^{i\nu}\partial_{i}\partial_{\alpha}F^{\alpha 0} - \partial_{0}\partial_{\alpha}F^{\alpha\nu}\right) + \frac{1}{\xi }\eta ^{0\nu }\partial _{\mu }A^{\mu }+\frac{2}{\xi m_{p}^{2}}
\eta ^{i\nu }\partial _{i}\partial _{\mu }A^{\mu }+\frac{1}{\xi m_{p}^{2}}
\eta ^{0\nu }\partial _{0}\partial _{\mu }A^{\mu }, \\
\Phi^{\nu} &= \frac{1}{m_{p}^{2}} \left(\partial _{\alpha
}F^{\alpha \nu }-\eta ^{\nu 0}\partial _{\alpha }F^{\alpha 0}\right)+\frac{1}{\xi
m_{p}^{2}}\eta ^{0\nu }\mathcal{\partial }^{0}\partial _{\nu }A^{\nu },
\end{align}
from which, together with the relations
\begin{align}
F^{\mu\nu}F_{\mu\nu} &= F^{kj}F_{kj} + 2\left(\Gamma _{j}-\partial _{j}A^{0}\right)^{2}, \\
\partial^{\mu}F_{\mu\beta}\partial_{\alpha}F^{\alpha\beta} &= - \frac{1}{2\xi}\left(\Gamma ^{0}\Gamma ^{0}+\partial_{i}A^{i}\partial _{j}A^{j}\right) + \frac{1}{2\xi m_{p}^{2}}\left(\partial _{\mu }\Gamma
^{\mu }\partial ^{\mu }\Gamma _{\mu }+\partial _{i}\partial
_{j}A^{j}\partial ^{i}\partial _{k}A^{k}\right), \\
\partial_{\mu}\Gamma^{\mu} &= \xi m_{p}^{2}\Phi ^{0}, \\
\partial _{0}\Gamma ^{k} &= \partial ^{k}\Gamma _{0} - \partial_{l}F^{lk}+m_{p}^{2}\Phi^{k},
\end{align}
it follows that the canonical Hamiltonian is written as
\begin{align}
H_{c} &= \int d^{3}x \ \bigg[ \Pi^{\nu}\Gamma_{\nu} - \Phi^{0}\partial_{i}\Gamma^{i} + \frac{\xi m_{p}^{2}}{2}\Phi^{0}\Phi^{0} + \Phi_{k}\left(\partial^{k}\Gamma_{0} - \partial_{l}F^{lk} + \frac{m_{p}^{2}}{2}\Phi^{k}\right) + \frac{1}{4}F^{kj}F_{kj} \nonumber\\
&\quad+\frac{1}{2}\left(\Gamma _{j} - \partial _{j}A_{0}\right)^{2} + \frac{1}{2m_{p}^{2}}\left(\partial _{j}\partial ^{j}A^{0} - \partial _{j}\Gamma ^{j}\right)^{2} + \frac{1}{2\xi}\left(\Gamma ^{0}\Gamma ^{0}+\partial _{i}A^{i}\partial _{j}A^{j}\right) - \frac{1}{2\xi
m_{p}^{2}}\partial _{i}\partial _{j}A^{j}\partial ^{i}\partial _{k}A^{k} \bigg].%
\label{CanonicalHamiltonianVectorSector}
\end{align}
Having determined $H_c$, the quantization of this sector is straightforward, namely, the construction of the quantum dynamics in the Hilbert space is given by the Dirac correspondence principle
\begin{align}
\left[\hat{A}^{\mu}(x),\hat{\Pi}_{\nu}(y)\right] &=i\delta _{\nu }^{\mu }\delta
^{3}(\vec{x}-\vec{y}), \nonumber\\
\left[\hat{\Gamma}^{\mu}(x),\hat{\Phi}_{\nu }(y)\right] &= i\delta _{\nu }^{\mu
}\delta ^{3}(\vec{x}-\vec{y}).%
\label{quantic2}
\end{align}

The description of this sector in thermodynamic equilibrium is obtained by introducing the density matrix of states

\begin{equation}
\hat{\rho_{s}}(\beta) = \exp \left(-\beta \hat{H}\right),
\label{DensityMatrix}
\end{equation}
where $\hat{H}$ is (\ref{CanonicalHamiltonianVectorSector}) with the fields turned to operators.
Applying the similarity transformation $\hat{\mathcal{O}^{s}}= \hat{\rho_{s}}^{-1}\hat{\mathcal{O}}\hat{\rho_{s}}$ in (\ref{quantic2})
we obtain the commutators in thermodynamic equilibrium

\begin{align}
\left[ \hat{A}_{\mu}^{s}\left(\vec{x},\tau\right),\hat{\Pi}_{\nu }^{s}\left(\vec{y},\tau\right)\right] &= i\delta _{\mu \nu }\delta ^{3}\left(\vec{x}-\vec{y}\right)\\
\left[ \hat{\Gamma}_{\mu }^{s}\left(\vec{x},\tau \right),\hat{\Phi}_{\nu }^{s}\left(\vec{y}
,\tau \right)\right] &= i\delta _{\mu \nu }\delta ^{3}\left(\vec{x}-\vec{y}\right).
\end{align}
With that we have constructed the density matrix of states $\hat{\rho}_{s}$ and the Hamiltonian that would describe the covariant thermal quantum equations of the radiation sector in the presence of external sources. Hence, we can proceed as in the previous subsection and perform the corresponding computations. However, it is more elegant to use the Nakanishi auxiliary field method \cite{Nakani} together with the Schwinger quantum action principle to find the partition function. 

Considering that the structures presented in the quantum description of fields in thermodynamic equilibrium are structures in the Euclidean space, the Lagrangian density which describes the radiation sector with interaction and sources is given by
\begin{align}
\hat{\mathcal{L}}_{N} &= \frac{1}{4}\hat{\mathcal{F}^{s}}_{\mu \nu }\hat{\mathcal{F}^{s}}_{\mu\nu} + \frac{1}{2m_{p}}\mathcal{\partial }_{\mu }\hat{\mathcal{F}^{s}}_{\mu \lambda }\mathcal{\partial}_{\theta }\hat{\mathcal{F}^{s}}_{\theta\lambda} + \frac{1}{2}\left\{\hat{B^{s}},G\left[\hat{A}^s\right]\right\} + \frac{\xi}{2}{\hat{B^{s}}}^{2} - ie\hat{A}_{\mu }^{E}\hat{\overline{\psi^s}}\beta _{\mu }^{E}\hat{\psi^{s}}  - \mathcal{J}_{\mu}\hat{A}_{\mu }^{s}, \\
\mathcal{\hat{F}}^{s}_{\mu \nu } &= \partial _{\mu }\hat{A}^{s}_{\nu} - \partial _{\nu }\hat{A}^{s}_{\mu } \nonumber,
\end{align}
where $\hat{B}^{s}$ is the Nakanishi-Lautrup auxiliary field and $G\left[\hat{A}^s\right]$ is the operator of gauge condition. To find the equations of motion in thermodynamic equilibrium we will use the Schwinger variational principle which states that
\begin{equation}
\delta\left( \hat{\psi}^{s}(t_1),\hat{\phi}^{s}(t_2) \right) = \frac{i}{\hbar} \left( \hat{\psi}^{s}(t_1), \delta \int_{t_2}^{t_1} d\tau \hat{\mathcal{L}}^{s}_N(\tau) \hat{\phi}^{s}(t_2) \right),
\label{eq:SchwingerTheorem}
\end{equation}
where $\delta$ is the gauge variation
\begin{equation}
\hat{A}^{s}_{\mu }\rightarrow \hat{A}^{s}_{\mu }+\delta \hat{A}^{s}_{\mu }, \qquad
\hat{B}^{s}\rightarrow \hat{B}^{s}+\delta \hat{B}^{s}.
\end{equation}%
After imposing $\delta \hat{S}^{s} = \hat{0}$, we should find the equations of motion arising from standard canonical quantization. The no-mixing gauge choice is written as

\begin{equation}
G\left[\hat{A}^s\right] = \left( \frac{\Delta }{m_{p}^{2}} + 1 \right)^{1/2}\partial_{\mu}\hat{A}_{\mu}^s.
\end{equation}
from which, together with the use of the relations
\begin{align}
\frac{1}{4}\hat{\mathcal{F}}^{s}_{\mu \nu }\hat{\mathcal{F}}^{s}_{\mu \nu } &\longrightarrow -\frac{1}{2}\hat{A}^{s}_{\mu} \left( \delta _{\mu \nu }\Delta + \partial_{\mu}\partial _{\nu } \right) \hat{A}^{s}_{\nu }, \\
\frac{1}{2m_{p}^{2}}\mathcal{\partial }_{\mu }\hat{\mathcal{F}}^{s}_{\mu \lambda
}\mathcal{\partial }_{\theta }\hat{\mathcal{F}}^{s}_{\theta \lambda } &\longrightarrow -\frac{1}{2
}\hat{A}^{s}_{\mu } \left( \delta _{\mu \nu }\Delta +\partial _{\mu }\partial
_{\nu } \right)\frac{\Delta }{m_{p}^{2}}\hat{A}^{s}_{\nu },
\end{align}
it follows that the equations of motion are

\begin{align}
- \left( \frac{\Delta }{m_{p}^{2}}+1 \right)\left(\delta _{\mu \nu }\Delta +\partial _{\mu
}\partial _{\nu }\right)\hat{A}^{s}_{\nu }-\left(\frac{\Delta }{m_{p}^{2}}+1\right)^{1/2}\partial _{\mu }\hat{B}^{s} - ie\hat{\overline{\psi^s}}\beta_{\mu }^{E}\hat{\psi^{s}} - \mathcal{J}_{\mu} &= 0, \\
- \frac{1}{\xi}\left(\frac{\Delta }{m_{p}^{2}}+1\right)^{1/2}\partial
_{\nu }\hat{A}^{s}_{\nu } &= B^{s},
\end{align}
or, equivalently,

\begin{equation}
- \left( \frac{\Delta ^{2}}{m_{p}^{2}}+1 \right) \left[\delta _{\mu \nu }\Delta + \left(1 - \frac{1}{\xi }\right)\partial _{\mu }\partial _{\nu }\right]\hat{A}^{s}_{\nu } = ie\hat{\overline{\psi^s}}\beta _{\mu }^{s}\hat{\psi^{s}} + \mathcal{J}_{\mu }.
\label{ftemp}
\end{equation}%
We immediately notice that in order to solve for $A_{\nu}$ we must know how to deal with the the Podolsky differential operator

\begin{equation}
P_{\mu\nu}^{\left(m_{p}^{2},\alpha \right)} := -\left(\frac{\Delta }{m_{p}^{2}} + 1\right)\left[\delta _{\mu \nu }\Delta +\left(1-\frac{1}{\xi }\right)\partial _{\mu }\partial_{\nu }\right].
\end{equation}
Something interesting about this analysis is that we have extended two methods to finite temperature: the Nakanishi method to fix the physical degrees of freedom in a covariant way by Lagrange multiplier and the Schwinger variational method to find the equations of motion.

As the last step we obtain the partition function that describes the vector sector with interaction in the no-mixing gauge. The density matrix of states is given again by (\ref{DensityMatrix}) with
\begin{align}
\hat{H} &=\int d^{3}z \bigg[ \hat{\Pi}^{\nu} \hat{\Gamma}_{\nu} - \hat{\Phi}^{0}\partial_{i}\hat{\Gamma}^{i} + \frac{\xi m_{p}^{2}}{2} \hat{\Phi}^{0}\hat{\Phi}^{0} + \hat{\Phi}_{k} \left( \partial^{k}\hat{\Gamma}_{0} - \partial_{l}\hat{F}^{lk} + \frac{m_{p}^{2}}{2}\hat{\Phi}^{k} \right) + \frac{1}{4}\hat{F}^{kj}\hat{F}_{kj} + \frac{1}{2} \left(\hat{\Gamma}_{j} - \partial_{j}\hat{A}_{0}\right)^{2} \nonumber\\
&\quad+\frac{1}{2m_{p}^{2}} \left( \partial _{j}\partial ^{j}\hat{A}^{0}-\partial _{j}\hat{\Gamma}
^{j}\right)^{2} + \frac{1}{2\xi }\left( \hat{\Gamma} ^{0}\hat{\Gamma} ^{0}+\partial _{i}\hat{A}^{i}\partial
_{j}\hat{A}^{j}\right) - \frac{1}{2\xi m_{p}^{2}}\partial _{i}\partial _{j}\hat{A}^{j}\partial
^{i}\partial _{k}\hat{A}^{k}+e\hat{A}_{\mu}\hat{\overline{\psi}}\beta^{\mu }\hat{\psi} + J^{\mu}\hat{A}_{\mu } \bigg].%
\end{align}%
Note that the ensemble in question is the canonical ensemble because we have no charge conservation. In this case the equation (\ref{ftemp}) can be written as follows
\begin{equation}
P_{\mu\nu}^{\left(m_{p}^{2},\alpha \right)} \frac{\delta\hat{\rho_{s}}(\beta)}{\delta\mathcal{J}^{\nu }\left(\vec{x},\tau\right)} = ie\hat{\overline{\psi^s}}\beta_{\mu}^{E}\hat{\psi^{s}} + \mathcal{J}_{\mu}\hat{\rho_{s}}(\beta ).
\label{PmunuZ2}
\end{equation}%
After taking the trace we have a functional equation for the partition function

\begin{equation} 
P_{\mu\nu}^{\left(m_{p}^{2},\alpha \right)} \frac{\delta Z_{2}[\mathcal{J}]}{\delta
\mathcal{J}^{\nu }(\vec{x},\tau )} = ie\overline{\psi^s}\beta_{\mu}^{E}\psi^{s} + \mathcal{J}_{\mu }Z_{2}\left[\mathcal{J},\overline{\eta},\eta \right].
\end{equation}
We look for a functional Fourier solution of the previous equation as follows
\begin{equation}
Z_{2}[\mathcal{J}] = \int \mathcal{D} \mathcal{A} \ \widetilde{Z}_{2} \left[ \mathcal{A},\overline{\psi},\psi\right]\exp \left(\mathcal{J}_{\mu }\mathcal{A}_{\mu }\right),
\end{equation}
from which (\ref{PmunuZ2}) takes the form

\begin{equation}
\int \mathcal{D} \mathcal{A} \left( P_{\mu\nu}^{\left(m_{p}^{2},\alpha \right)} A_{\nu} - ie\overline{\psi}\beta_{\mu}^{E}\psi \right) \widetilde{Z}_{2} \left[ \mathcal{A},\overline{\psi},\psi \right] \exp \left(%
\mathcal{J}_{\mu }\mathcal{A}_{\mu } \right) = \mathcal{J}_{\mu }Z_{2}[\mathcal{J}].
\end{equation}%
Defining
\begin{equation}
S_{2} := \int d^{4}x \ \left[ \frac{1}{2}\mathcal{A}_{\mu } P_{\mu\nu}^{\left(m_{p}^{2},\alpha \right)} \mathcal{A}_{\nu} - ie\mathcal{A}_{\mu }\overline{\psi}\beta _{\mu }^{E}\psi \right],
\end{equation}
such that
\begin{equation*}
\frac{\delta S_{2}}{\delta \mathcal{A}_{\mu }} = P_{\mu\nu}^{\left(m_{p}^{2},\alpha \right)} \mathcal{A}_{\nu } - ie\overline{\psi}\beta _{\mu }^{E}\psi,
\end{equation*}%
we realize that the partition function that describes the vector sector is
\begin{equation}
Z_{2}[\mathcal{J}] = \int \mathcal{D} \mathcal{A} \ \exp \left(-S_{2}+\mathcal{J}_{\mu }\mathcal{%
A}_{\mu }\right).
\end{equation}

\subsection{The ghost sector}
The Lagrangian density which describes the ghost sector in the no-mixing gauge is given by \cite{And}
\begin{equation}
\mathcal{L} = i\overline{c} \left( \frac{\Delta}{m_{p}^{2}} + 1\right) \Delta c - \frac{1}{2}\phi \left( \frac{\Delta}{m_{p}^{2}} + 1\right)\phi.
\end{equation}
From there, we are able to build all the structure of the ghost sector. Classically, the action that describes the ghost sector, that comes from a study in the \textit{no-mixing} gauge, can be defined as
\begin{equation}
\mathcal{S} = \int d^{4}x \  \left[
i\partial_{\mu}{\overline{c}}\partial^{\mu}{c} - \frac{i}{m_{p}^{2}}\square{\overline{c}}\square{c} - \frac{1}{2m_{p}^{2}}\partial_{\mu}\phi \partial^{\mu }\phi + \frac{1}{2}\phi^{2} \right],
\label{ghost}
\end{equation}%
wherein $c, \overline{c}$ are Grassmanian fields and $\phi$ a real field. In this case, the classical equations of motion are given by
\begin{align}
\left( 1 + \frac{\square}{m_{p}^{2}} \right)\square{c} &= 0, \nonumber \\
\overline{c} \left( 1 + \frac{\overleftarrow{\square}}{m_{p}^{2}} \right)\overleftarrow{\square} &= 0, \label{eqmov} \\
\left( \frac{\square}{m_{p}^{2}} + 1 \right)\phi &= 0 \nonumber .%
\end{align}
On the other hand, the canonical Hamiltonian density is given by
\begin{equation}
\mathcal{H} = \left( \partial_{0}\overline{c} \right)\pi + \left( \partial_{0}\partial_{0}\overline{c} \right)P + \overline{\pi} \left(\partial_{0}c \right) + \overline{P} \left(\partial _{0}\partial _{0}c \right) + p \left(\partial_{0}\phi\right) - \mathcal{L},
\end{equation}%
where the canonical momenta have been defined as follows

\begin{align}
\pi &:= \frac{\partial\mathcal{L}}{\partial\left(\partial_{0}{\overline{c}}\right)} - 2\partial_{i}\left(\frac{\partial\mathcal{L}}{\partial\left(\partial_{i}\partial_{0}{\overline{c}}\right)}\right) - \partial_{0}\left(\frac{\partial\mathcal{L}}{\partial\left(\partial_{0}\partial_{0}{\overline{c}}\right)}\right)  = i \left( 1 + \frac{\square}{m_{p}^{2}} \right)\partial_{0}{c} \nonumber,\\
\overline{\pi} &:= \frac{\partial\mathcal{L}}{\partial\left(\partial_{0}{c}\right)} - 2\partial_{i}\left(\frac{\partial\mathcal{L}}{\partial\left(\partial_{i}\partial_{0}{c}\right)}\right) - \partial_{0}\left(\frac{\partial\mathcal{L}}{\partial\left(\partial_{0}\partial_{0}{c}\right)}\right)  = -i \left( 1 + \frac{\square}{m_{p}^{2}} \right)\partial_{0}{\overline{c}} \nonumber, \\
P &:= \frac{\partial\mathcal{L}}{\partial\left(\partial_{0}\partial_{0}{\overline{c}}\right)} = -i\frac{\square}{m_{p}^{2}}{c}, \\
\overline{P} &:= \frac{\partial\mathcal{L}}{\partial\left(\partial_{0}\partial_{0}{c}\right)} = i\frac{\square}{m_{p}^{2}}{\overline{c}} \nonumber, \\
p &:= \frac{\partial\mathcal{L}}{\partial\left(\partial_{0}\phi\right)} = -\frac{1}{m_{p}^{2}}\partial_{0}\phi \nonumber.%
\label{CanoMomentaGhost}
\end{align}
Moreover, the action (\ref{ghost}) is invariant by the transformations
\begin{align}
{\overline{c}}\rightarrow {\overline{c}} + \delta {\overline{c}}, \qquad \delta{\overline{c}} = -i\delta \theta {\overline{c}}, \\
c\rightarrow c+\delta c, \qquad \delta c = i\delta \theta c.
\end{align}%
which implies the existence of a ghost charge via Noether theorem, namely,
%

\begin{equation}
iQ_{0} = i\overline{c}  \left( 1 + \frac{\square}{m_{p}^{2}} \right)\partial_{0}c - i\left( 1 + \frac{\square}{m_{p}^{2}} \right)\left(\partial_{0}\overline{c}\right)c + \frac{i}{m_{p}^{2}}\square\overline{c} \left( \partial_{0}c \right) - \frac{i}{m_{p}^{2}}\left(\partial_{0}\overline{c}\right)\square c.
\label{eq:ChargeClassicalResultGhost}
\end{equation}

The quantization of the ghost sector is perform by first writing the ghost Hamiltonian density explicitly



\begin{equation}
\mathcal{H} = \overline{D}\pi - \left(\partial_{k}\partial^{k}\overline{c}\right)P + \overline{\pi}D - \overline{P}\left(\partial_{k}\partial^{k}c\right) + i\left(\overline{D}D + \partial_{k}\overline{c}\partial^{k}{c}\right)\\
-im_{p}^{2}\overline{P}P + \frac{1}{2}{m}_{p}^{2}{p}^{2} + \frac{1}{2m_{p}^{2}}\partial_{k}\phi\partial^{k}\phi - \frac{1}{2}\phi^{2},%
\label{eq:HamiltonianClassicalResultGhost}
\end{equation}
with the definition $D=\partial_{0}c$. As a result, the Hamilton equations are obtained by straightforward computations, i.e.,

\begin{align}
\dot{c} &= \frac{\delta\mathcal{H}}{\delta\overline{\pi}},\qquad 
\dot{\pi} = -\frac{\delta\mathcal{H}}{\delta\overline{c}}, \qquad
\overset{.}{\overline{c}} = \frac{\delta\mathcal{H}}{\delta\pi}, \qquad
\overset{.}{\overline{\pi}} = -\frac{\delta\mathcal{H}}{\delta {c}}, \qquad
\dot{D} = \frac{\delta\mathcal{H}}{\delta\overline{P}}, \qquad \nonumber\\
\dot{P} &= -\frac{\delta\mathcal{H}}{\delta\overline{D}}, \qquad
\overset{.}{\overline{D}} = \frac{\delta\mathcal{H}}{\delta P}, \qquad
\overset{.}{\overline{P}} = -\frac{\delta\mathcal{H}}{\delta {D}}, \qquad
\dot{\phi} = \frac{\delta\mathcal{H}}{\delta p}, \qquad
\dot{p} = -\frac{\delta\mathcal{H}}{\delta\phi}.%
\end{align}
However, since we are dealing with real and grassmann variables we have to introduce an extension of the Poisson brackets known as Berezin brackets. In view of their definition we can write the fundamental Berezin brackets as 
\begin{align}
\left\{ c(x),\overline{\pi}(y) \right\}_{B} &= \left\{ \overline{\pi}(x),c{(y)} \right\}_{B} = \delta^{3} \left( \vec{x}-\vec{y} \right), \qquad \left\{ \overline{D}(x),P(y) \right\}_{B} = \left\{P(x),\overline{D}(y) \right\}_{B} = \delta^{3} \left(\vec{x}-\vec{y}\right), \nonumber\\
\nonumber\\
\left\{ \overline{c}(x),\pi(y) \right\}_{B} &= \left\{\pi(x),\overline{c}(y) \right\}_{B} = \delta^{3} \left( \vec{x}-\vec{y} \right), \qquad \left\{ \phi(x),p(y) \right\}_{B} = -\left\{p(x),\phi(y) \right\}_{B} = \delta^{3} \left(\vec{x}-\vec{y}\right), \nonumber\\
\nonumber\\
\left\{ D(x),\overline{P}(y) \right\}_{B} &= \left\{ \overline{P}(x),D(y) \right\}_{B} = \delta^{3} \left( \vec{x}-\vec{y} \right).
\end{align}%
which become graded commutators $[\cdot,\cdot]_{\mathfrak{g}}$ in the quantum theory.

As we are interested in studying fields in thermodynamic equilibrium, the previous discussion leads us to define the density matrix of states in the grand-canonical ensemble of the ghost sector with sources in the following way
\begin{equation}
\hat{\rho_{gs}}(\beta) = \exp \left[ -\beta \left( \hat{H} - \mu_{g}\hat{Q} \right) \right] \\
\end{equation}
wherein the Hamiltonian and the charge are given by (\ref{eq:HamiltonianClassicalResultGhost}) and (\ref{eq:ChargeClassicalResultGhost}), respectively, after the corresponding transition to operators is made.

Applying a similarity transformation $\hat{\mathcal{O}^{gs}} \left(\vec{x},\tau \right) = \hat{\rho_{gs}}^{-1}(\tau)\hat{\mathcal{O}}\left(\vec{x}\right)\hat{\rho_{gs}}(\tau)$ in the stationary equations we obtain the fundamental commutation relations in thermodynamic equilibrium

\begin{align}
\left[ \hat{c}^{gs} \left(\vec{x},\tau\right),\hat{\overline{\pi}}^{gs}\left(\vec{y},\tau\right)\right]_{\mathfrak{g}} &= \left[\hat{\overline{\pi}}^{gs}\left(\vec{x},\tau\right),\hat{c}^{gs}\left(\vec{y},\tau\right)\right]_{\mathfrak{g}} = i\delta^{3}\left(\vec{x}-\vec{y}\right), \nonumber\\
\nonumber \\
\left[\hat{\overline{c}}^{gs}\left(\vec{x},\tau\right),\hat{\pi}\left(\vec{y},\tau\right)\right]_{\mathfrak{g}} &= \left[\hat{\pi}^{gs}\left(\vec{x},\tau\right),\hat{\overline{c}}^{gs}\left(\vec{y},\tau\right)\right]_{\mathfrak{g}} = i\delta ^{3}\left(\vec{x}-\vec{y} \right), \nonumber\\
\nonumber \\
\left[\hat{D}^{gs}\left(\vec{x},\tau\right),\hat{\overline{P}}^{gs}\left(\vec{y},\tau\right) \right]_{\mathfrak{g}} &= \left[ \hat{\overline{P}}^{gs}\left(\vec{x},\tau\right),\hat{D}^{gs}\left(\vec{y},\tau\right)\right]_{\mathfrak{g}} = i\delta^{3} \left(\vec{x}-\vec{y}\right),\\
\nonumber\\
\left[\hat{\overline{D}}^{gs}\left(\vec{x},\tau\right),\hat{P}^{gs}\left(\vec{y},\tau\right)\right]_{\mathfrak{g}} &= \left[\hat{P}^{gs}\left(\vec{x},\tau\right),\hat{\overline{D}}^{gs}\left(\vec{y},\tau\right)\right]_{\mathfrak{g}} = i\delta ^{3}\left(\vec{x}-\vec{y}\right), \nonumber\\
\nonumber\\
\left[\hat{\phi}^{gs}\left(\vec{x},\tau\right),\hat{p}^{gs}\left(\vec{y},\tau\right)\right]_{\mathfrak{g}} &= -\left[\hat{p}^{gs}\left(\vec{x},\tau \right),\hat{\phi}^{gs}\left(\vec{y},\tau\right)\right]_{\mathfrak{g}} = i\delta^{3} \left(\vec{x}-\vec{y}\right) \nonumber.%
\end{align}

The next step is to determine the equations of motion in thermodynamic equilibrium. Firstly, for $\hat{c}$ we get
\begin{align}
\frac{\partial}{\partial\tau}\left(\hat{c}^{gs}\right) &= \frac{\partial}{\partial\tau}\bigg( e^{\tau \left(\hat{H} - \mu_{g}\hat{Q}\right)}\hat{c} \ e^{-\tau \left(\hat{H} - \mu_{g}\hat{Q}\right)} \bigg), \label{FirstDerivativeTau}\\
&=e^{\tau \left(\hat{H} - \mu_{g}\hat{Q}\right)} \left[\hat{H} - \mu_{g}\hat{Q}, \hat{c}\right]e^{-\tau \left(\hat{H} - \mu_{g}\hat{Q}\right)}.  \nonumber
\end{align}
Using the identity $\left[\hat{A}\hat{B},\hat{C}\right] = \hat{A} \left\{\hat{B},\hat{C}\right\} - \left\{\hat{A},\hat{C}\right\}\hat{B}$, we finally obtain

\begin{equation}
\frac{\partial}{\partial\tau}\left(\hat{c}^{gs}\right) = -i\hat{D}^{gs}\left(\vec{x},\tau\right) + \mu_{g}\hat{c}^{gs}\left(\vec{x},\tau\right).%
\label{Itera}
\end{equation}%
%
Taking the second derivative and using (\ref{Itera}) again we find
\begin{equation}
\frac{\partial^2\hat{c}^{gs}}{\partial\tau^2} = -i\frac{\partial \hat{D}^{gs}\left(\vec{x},\tau\right)}{\partial \tau } + \mu _{g}\left[-i\hat{D}^{gs}\left(\vec{x},\tau \right) + \mu _{g}\hat{c}^{gs}\left(\vec{x},\tau \right)\right]. 
\end{equation}
For the first term we can use an analogous equation to (\ref{FirstDerivativeTau}), i.e,
\begin{align}
\frac{\partial \hat{D}^{gs}\left(\vec{x},\tau \right)}{\partial \tau } &= e^{\tau \left(\hat{H} - \mu_{g}\hat{Q}\right)} \left[\hat{H} - \mu_{g}\hat{Q}, \hat{D}\right]e^{-\tau \left(\hat{H} - \mu_{g}\hat{Q}\right)}, \\
\left[\hat{H} - \mu_{g}\hat{Q}, \hat{D}\right] &= \int d^{3}y \ \left[ -\hat{\overline{P}} \left( \partial_{k}\partial^{k}\hat{c}\right) - im_{p}^{2}\hat{\overline{P}}\hat{P} - i\mu_{g}\hat{\overline{P}}\hat{D},\hat{D}\left(\vec{x}\right) \right]_{\mathfrak{g}},  \\
&= i\partial_{k}\partial^{k}\hat{c} - m_{p}^{2}\hat{P} - \mu_{g}\hat{D}. %
\end{align}%
Therefore,

%
%
\begin{equation}
\hat{P}^{gs}\left(\vec{x},\tau\right) =
\left( i\frac{\Delta}{m_{p}^{2}} + i\frac{\mu_{g}^{2}}{m_{p}^{2}} \right)\hat{c}^{gs}\left(\vec{x},\tau\right).
\end{equation}
Taking another derivative with respect to $\tau$ and performing a similar computation for the left hand side, in which the graded commutator $\left[\hat{H} - \mu_{g}\hat{Q}, \hat{P}\right]_{\mathfrak{g}}$ appears, we obtain the equation

\begin{equation}
\hat{\pi}^{gs}\left(\vec{x},\tau\right) = 
-i\left( 1 - \frac{\Delta}{m_{p}^{2}}  - \frac{\mu_{g}^{2}}{m_{p}^{2}} \right) \hat{D}^{gs}\left(\vec{x},\tau\right).%
\end{equation}
Thus, following the same reasoning, we take an additional derivative and compare with $\left[\hat{H}-\mu_g\hat{Q},\hat{\pi}\right]_{\mathfrak{g}}$. We will end up with the equation of motion for the ghost field in thermodynamic equilibrium (similarly for $\hat{\overline{c}}^s$)

\begin{equation}
-i \left( 1 - \frac{\Delta}{m_p^2} \right) \Delta \hat{c}^{gs}\left(\vec{x},\tau\right) = \zeta
\left(\vec{x},\tau\right), \qquad \mu_{g} = 0.
\label{eq:ghost}
\end{equation}
In the same way, we find the equation of motion for $\hat{\phi}$. Again,
\begin{align}
\frac{\partial\hat{\phi}^{gs}\left(\vec{x},\tau\right)}{\partial\tau} &= \exp\left[\tau(\hat{H} - \mu_{g}\hat{Q})\right]\left[\hat{H} - \mu_{g}\hat{Q},\hat{\phi}\right]_{\mathfrak{g}} \exp\left[-\tau(\hat{H} - \mu_{g}\hat{Q})\right], \\
\left[\hat{H} - \mu_{g}\hat{Q},\phi\left(\vec{x}\right)\right] &= \left[ \int d^{3}y \  \frac{1}{2}{m}_{p}^{2}p^2,\hat{\phi}\left(\vec{x}\right)\right]= -im_{p}^{2}\hat{p},
\end{align}
or

\begin{equation}
	\frac{\partial\hat{\phi}^{gs}\left(\vec{x},\tau\right)}{\partial\tau} = -im^2_p\hat{p}^{gs}.
\end{equation}
Taking an additional derivative and computing $\left[\hat{H}-\mu_g\hat{Q},\hat{p}\right]_{\mathfrak{g}}$ we finally end up with

\begin{equation}
\left(1 - \frac{\Delta}{m_{p}^{2}}\right)\hat{\phi}^{gs}\left(\vec{x},\tau\right) = -J\left(\vec{x},\tau\right).
\label{eq:phighost}
\end{equation}


As we can see, there is a relation between the classical equations (\ref{eqmov}) arising from the action (\ref{ghost}) and the quantum equations in thermodynamic equilibrium, namely,

\begin{equation}
\square ~ \longleftrightarrow ~ -\Delta .
\end{equation}%
In order for the quantum equations in thermodynamic equilibrium to be properly defined we should use the following classical action
\begin{equation}
\begin{array}{l}
\mathcal{S}=\int d^{4}x\mathcal{L} \\
\\
\mathcal{L}:=i{\bar{c}(1-}\frac{\square }{m_{p}^{2}})\square {c}+\frac{1}{2}%
\phi (1-\frac{\square }{m_{p}^{2}})\phi
\end{array}%
\end{equation}%
Therefore, the quantum equations in thermodynamic equilibrium would be
\begin{equation}
\begin{array}{l}
i(1+\frac{\Delta }{m_{p}^{2}})\Delta \widehat{{c}}^{gs}(\vec{x},\tau )=\zeta (%
\vec{x},\tau ),\text{ \ \ }\mu _{g}=0 \\
\\
-i(1+\frac{\Delta }{m_{p}^{2}})\Delta \widehat{{{\bar{c}}}}^{gs}(\vec{x},\tau
)=\bar{\zeta}(\vec{x},\tau ),\text{ \ \ }\mu _{g}=0 \\
\\
(1+\frac{\Delta }{m_{p}^{2}})\widehat{\phi }^{gs}(\vec{x},\tau )=-J(\vec{x}%
,\tau ).%
\end{array}
\end{equation}
In order to find the partition function, we rewrite the previous equations of motion as follows
\begin{align}
i\left( 1 + \frac{\Delta}{m_{p}^{2}} \right) \Delta \frac{\delta\hat{\rho}_{gs}(\beta)}{\delta\overline{\zeta}\left(\vec{x},\tau\right)} &= \zeta\left(\vec{x},\tau\right)\hat{\rho}_{gs}(\beta) \\
-i\left(1 + \frac{\Delta}{m_{p}^{2}}\right) \Delta \frac{\delta\hat{\rho}_{gs}(\beta)}{\delta\zeta\left(\vec{x},\tau\right)} &= \overline{\zeta}\left(\vec{x},\tau\right)\hat{\rho}_{gs}(\beta) \\
\left(1 + \frac{\Delta}{m_{p}^{2}}\right) \frac{\delta\hat{\rho}_{gs}(\beta)}{\delta J\left(\vec{x},\tau\right)} &= -J\left(\vec{x},\tau\right)\hat{\rho}_{gs}(\beta),
\end{align}%
after taking the trace we obtain

\begin{align}
i\left(1 + \frac{\Delta}{m_{p}^{2}}\right) \Delta \frac{\delta Z_{3}\left[\overline{\zeta},\zeta,J\right]}{\delta \overline{\zeta}\left(\vec{x},\tau\right)} &= \zeta\left(\vec{x},\tau\right)Z_{3}\left[\overline{\zeta},\zeta,J\right] \\
-i\left(1 + \frac{\Delta}{m_{p}^{2}}\right)\Delta \frac{\delta Z_{3}\left[\overline{\zeta},\zeta,J\right]}{\delta \zeta \left(\vec{x},\tau\right)} &= \overline{\zeta}\left(\vec{x},\tau\right)Z_{3}\left[\overline{\zeta},\zeta,J\right] \\
\left(1 + \frac{\Delta}{m_{p}^{2}}\right)\frac{\delta Z_{3}\left[\overline{\zeta},\zeta,J\right]}{\delta J\left(\vec{x},\tau\right)} &= -J\left(\vec{x},\tau\right)Z_{3}\left[\overline{\zeta},\zeta,J\right].%
\end{align}%
We will look for a solution in form of a functional Fourier transform of the previous equation as follows
\begin{equation}
Z_{3}\left[\overline{\zeta},\zeta,J\right] = \int \mathcal{D}\overline{c} \mathcal{D}{c} \mathcal{D}\phi \ \widetilde{Z}_{3}\left[\overline{c},c,\phi\right]\exp\left(\overline{c}\zeta + \overline{\zeta}c + J\phi \right).
\end{equation}%
Defining

\begin{equation}
S_{3} = \int d^{4}x \ \left[i\overline{c}\left(\frac{\Delta }{m_{p}^{2}} + 1\right)\Delta c - \frac{1}{2}\phi \left(\frac{\Delta }{m_{p}^{2}}+1\right)\phi \right]
\end{equation}%
we notice that the partition function that describes the ghost sector is given by
\begin{equation}
Z_{3}\left[\overline{\zeta},\zeta,J\right] = \int \mathcal{D}\overline{c} \mathcal{D}{c} \mathcal{D}\phi \ \exp \left(-S_{3}+\bar{c}%
\zeta +\bar{\zeta}c+J\phi \right).
\end{equation}

\section{Partition function and physical degrees of freedom}\label{sec3}

As we know, the second law of thermodynamics is based on the Kelvin-Clausius prescription, using Carnot's cycles and Clausius's theorem to formulate a state function called entropy. The latter was formulated axiomatically by Carathéodory. However, when we construct the thermodynamics of quantum fields we use as an ontological starting point the Gibbs-Helmohtz formalism which is given in terms of a variational principle of maximum entropy. Accordingly, this variational formulation was put on a logical basis by Tisza-Callen \cite{Callen, LTisza} which equips us with a pleasant reading formalism that allows us to respond conceptually how entropy translates into thermal quantum fields and also to study the state equations that describe the interactions.

In view of this, we can say that the density matrix describing the grand-canonical ensemble of GSDKP in thermodynamic equilibrium in a covariant dynamics is written explicitly by defining the Gibbs entropy

\begin{equation}
\hat{S}_{\text{entropy}} = \left\langle -\ln \left[\hat{\rho}_{gs}(\beta)\right] \right\rangle,
\end{equation}%
with 
\begin{equation}
	\hat{\rho}_{gs}(\beta) = \exp\left[ -\beta\left(\hat{\mathbb{H}} - \mu_{e}\hat{N} - \mu_{g}\hat{Q}\right)\right], \qquad \hat{\mathbb{H}} = \hat{H}_{T} + \hat{H}_{g},
\end{equation}
wherein

\begin{align}
\hat{H}_{T} &= \int d^{3}x \ \bigg[ \hat{\Pi}^{\nu}\hat{\Gamma}_{\nu} - \hat{\Phi}^{0}\partial_{i}\hat{\Gamma}^{i} + \frac{\xi m_{p}^{2}}{2}\hat{\Phi}^{0}\hat{\Phi}^{0} + \hat{\Phi}_{k} \left(\partial^{k}\hat{\Gamma}_{0} - \partial_{l}\hat{F}^{lk} + \frac{m_{p}^{2}}{2}\hat{\Phi}^{k}\right) + \frac{1}{4}\hat{F}^{kj}\hat{F}_{kj} \nonumber\\
&\quad+\frac{1}{2}\left(\hat{\Gamma}_{j} - \partial_{j}\hat{A}_{0}\right)^{2} + \frac{1}{2m_{p}^{2}} \left(\partial_{j}\partial^{j}\hat{A}^{0} - \partial_{j}\hat{\Gamma}^{j}\right)^{2} + \frac{1}{2\xi}\left(\hat{\Gamma}^{0}\hat{\Gamma}^{0} + \partial_{i}\hat{A}^{i}\partial_{j}\hat{A}^{j}\right) - \frac{1}{2\xi m_{p}^{2}}\partial_{i}\partial_{j}\hat{A}^{j}\partial^{i}\partial_{k}\hat{A}^{k} \nonumber\\
&\quad-\frac{i}{2}\hat{\overline{\psi}}\beta^{i}\left(\overleftrightarrow{\partial_{i}}\hat{\psi}\right) + m\hat{\overline{\psi}}\hat{\psi} - e\hat{A}_{\mu}\hat{\overline{\psi}}\beta^{\mu}\hat{\psi} + J_{\mu}\hat{A}^{\mu} + \overline{\eta}\hat{\psi} + \hat{\overline{\psi}}\eta \bigg],\\
\hat{H}_{g} &= \int d^{3}x \ \bigg[ \hat{\overline{D}}\hat{\pi} - \left(\partial_{k}\partial^{k}\hat{c}\right)\hat{P} + \hat{\overline{\pi}}\hat{D} + \hat{\overline{P}}\left(-\partial_{k}\partial^{k}\hat{c}\right) + i\left(\hat{\overline{D}}\hat{D} + \partial_{k}\hat{\overline{c}}\partial^{k}\hat{c}\right) \nonumber \\
&\quad-im_{p}^{2}\hat{\overline{P}}\hat{P} + \frac{1}{2}{m}_{p}^{2}\hat{p}^{2} + \frac{1}{2m_{p}^{2}}\partial_{k}\hat{\phi}\partial^{k}\hat{\phi} - \frac{1}{2}\hat{\phi}^{2} + \overline{\zeta}\hat{c} + \hat{\overline{c}}\zeta + j\hat{\phi} \bigg], \\
i\hat{Q} &= \int d^{3}x \ \left[\hat{\overline{c}}\hat{\pi} + \hat{\overline{\pi}}\hat{c} + \hat{\overline{P}}\hat{D} + \hat{\overline{D}}\hat{P} \right], \\
\hat{N} &= \int d^{3}x \ \hat{\overline{\psi}}\beta^{0}\hat{\psi},%
\label{matrizdens}
\end{align}
and formulate the problem from a variational principle of entropy in such a way that

\begin{equation}
\delta \left[ \lambda \left\langle \hat{I} \right\rangle + \lambda_{U}\left\langle \hat{\mathbb{H}}\right\rangle + \lambda_{N}\left\langle \hat{N}\right\rangle + \lambda_{Q}\left\langle \hat{Q}\right\rangle - \left\langle -\ln \left[\hat{\rho}_{gs}(\beta)\right] \right\rangle \right] = 0.
\end{equation}

In order to obtain the Lagrange multipliers we compare the equation obtained from the extremization process with the thermodynamic equation that defines the grand-canonical potential $\Omega$,

\begin{equation}
\Omega =U -TS -\mu_{e}N -\mu _{g}Q,
\end{equation}
wherein $U=\left\langle \hat{\mathbb{H}}\right\rangle$ is the internal energy, $S=\left\langle\hat{S}\right\rangle$ is the entropy, $N=\left\langle\hat{N}\right\rangle$ is the number of charge particles and $Q=\left\langle\hat{Q}\right\rangle$ the number of ghost particles. Here it is important to pay attention that the instability problem was solved in Podosky theory following the Lagrangian achor methodology \cite{Russos} and in finite temperature there is no problem at all due to the fact that the internal energy of Podolsky theory is defined naturally as the sum of Maxwell and Proca (this fact reflex the existence of the Lagrangian anchor mechanism), seen in the Stephan-Boltzmann Law \cite{Boni}. So this instability do not affect the internal energy and others thermodynamic quantities.

From the equations of motion (\ref{eq:Scalar1}) and (\ref{seq}), (\ref{ftemp}), and (\ref{eq:ghost}) and (\ref{eq:phighost}) representing the scalar, vectorial and ghost sectors respectively of GSDKP in the non-mixing gauge we can construct the thermodynamic generator $Z\left[\mathcal{J},{\overline{\eta}},\eta,\overline{\zeta},\zeta,J\right]$ obtained as the solution of the functional equations
%


\begin{align}
\left[ \left( \beta_{\mu}^E \right)_{ab}\partial_{\mu}^{\mu_e} - m\delta_{ab} \right] \frac{\delta Z}{\delta\overline{\eta}_{b}} &= ie \left(\beta_{\mu}^E \right)_{ab}\frac{\delta^{2}Z}{\delta J_{\mu}\delta\overline{\eta}_{b}} + \eta Z \\
\frac{\delta Z}{\delta\eta_a} \left[ \left(\beta_{\mu}^E \right)_{ab}\overleftarrow{\partial_{\mu }^{-\mu_e}} + m\delta_{ab} \right] &= -ie \left(\beta_{\mu}^E\right)_{ab}\frac{\delta^{2}Z}{\delta J_{\mu}\delta\eta_{b}} + Z\overline{\eta} \\
P_{\mu\nu}^{\left(m_{p}^{2},\alpha\right)}\frac{\delta Z}{\delta \mathcal{J}_{\nu}} &= ie \left(\beta_{\mu }^{E}\right)_{ab}\frac{\delta Z_{GF}}{\delta \eta_{a}\delta\overline{\eta}_{b}} + \mathcal{J}_{\mu }Z \\
i \left( 1 + \frac{\Delta}{m_{p}^{2}} \right)\Delta\frac{\delta Z}{\delta\overline{\zeta}} &= \zeta Z \\
-i \left( 1 + \frac{\Delta}{m_{p}^{2}} \right)\Delta\frac{\delta Z}{\delta\zeta} &= Z\overline{\zeta} \\
\left( 1 + \frac{\Delta}{m_{p}^{2}} \right)\frac{\delta Z}{\delta J} &= JZ.
\end{align}
In this case $Z$ can be written as a functional Fourier transform
\begin{equation}
Z\left[\mathcal{J},{\overline{\eta}},\eta,\overline{\xi},\xi,J\right] = \int \mathcal{D}\mathcal{A} \mathcal{D}\overline{\psi} \mathcal{D}\psi \mathcal{D}{\overline{c}} \mathcal{D}c \mathcal{D}\phi \exp \left(-S_{T}^{\text{eff}} \right),
\end{equation}
with
\begin{align}
S_{T}^{\text{eff}} &= \int d^{4}x \ \bigg[ \frac{1}{2} \mathcal{A}_{\mu}P_{\mu\nu}^{\left(m_{p}^{2},\xi \right)}\mathcal{A}_{\nu} - \overline{\psi}\left(\beta_{\mu}^{E}D_{\mu}^{\left(e,\mu _{e}\right)} - m\right)\psi - i\overline{c} \left(\frac{\Delta}{m_{p}^{2}} + 1\right)\Delta c \nonumber\\
&\quad+\frac{1}{2}\phi \left(\frac{\Delta}{m_{p}^{2}} + 1\right)\phi + \mathcal{J}_{\mu}\mathcal{A}_{\mu} + \overline{\eta}\psi + \overline{\psi}\eta + \overline{\zeta}c + \overline{c}\zeta + J\phi \bigg].
\label{parteq}
\end{align}
It follows that in the free case, i.e., $e = 0$, we have
\begin{equation}
Z = Z_{\text{Podolsky}} Z_{\text{DKP}},
\end{equation}
where
\begin{align}
Z_{\text{Podolsky}} &= \det \left[ \triangle \right]^{-1} \det \left[ \triangle + {m^2_p}\right]^{-3/2}, \\
Z_{\text{DKP}} &= \det \left[ \beta_{\mu }^{E}\partial_{\mu }^{(\mu_{e})} - m\right]
\end{align}
It is worthwhile to emphasize that the ghosts sector eliminate the degrees of freedom of the vector sector maintaining the physical degrees of freedom of the theory. Note that in the no-mixing gauge choice the ghost has two sectors \cite{And}, one grassmanian and other scalar, and the scalar field eats the degrees of freedom of the grassmann field that eat the degrees of freedom of the vectorial field maintaining the physical degrees of freedom of Podolsky theory.

Therefore, we can write the grand-canonial potencial $\Omega$ as
\begin{equation}
	\Omega = -kT\ln Z = \Omega_{\text{Maxwell}} + \Omega_{\text{Proca}} + \Omega_{\text{DKP}}
\end{equation}
with
\begin{align}
\Omega_{\text{Maxwell}} &= \left(\frac{1}{2} + \frac{1}{2}\right) kT\sum_{n,\vec{p}} \ln\left[ \beta^{2} \left( \omega_{n}^{2} + \vec{p}^{2} \right) \right] \\
\Omega_{\text{Proca}} &= \left( \frac{1}{2} + \frac{1}{2} + \frac{1}{2} \right)
kT \sum_{n,\vec{p}}\ln\left[ \beta^{2} \left( \omega_{n}^{2} + \vec{p}^{2}  + m_p^2\right) \right] \\ 
\Omega_{\text{DKP}} &= \left( \frac{1}{2} + \frac{1}{2} \right)
kT\sum_{n,\vec{p}} \ln\left[ \beta^{2} \left( \left(\omega_{n}^{2} + i\mu_e\right)^2 + \vec{p}^{2}  + m_p^2\right) \right],
\end{align}
from which we clearly see the connection between the physical degrees of freedom and the equipartition theorem, five degrees of freedom for the Podolsky photons and two degrees of freedom for DKP scalars.

\section{Conclusion and final remarks}
\label{sec4}

In this work we have analyzed the link between the physical degrees of freedom, quantization and the description of thermodynamic equilibrium when studying the interaction between matter and radiation in the context of GSDKP.

Firstly, in constructing the transition amplitude, we saw the emergence of the constraints due to the construction of physical phase space and its importance in writing the measure of integration. Explicitly, in the analysis of the order of the equations, the physical degrees of freedom are closely related to the constraints. The phase space has 36 variables $(\Phi^{\nu },\Gamma _{\nu },\Pi^{\mu },A_{\mu };\mathcal{\bar{\psi}},
\mathcal{\psi },\bar{p},p)$ and 26 constraints $\left\{\chi ^{(1)},\bar{\chi}^{(1)},\chi ^{(2)},\bar{\chi}^{(2)};\varphi _{1},\varphi _{2},\varphi _{3},\Sigma _{1},\Sigma _{2},\Sigma_{3}\right\}$ so the physical phase space has 14 variables and 7 physical degrees of freedom. As we could see the quantization procedure by the Fadeev-Senjanovic analysis in the generalized Coulomb gauge and the Faddev-Popov-DeWitt procedure to get the covariant transition amplitude was not an easy task, due to the no-mixing gauge choice (pseudo differential structure) and the large amount of non-physical degrees of freedom. As in a popular saying, things need to get worse before they get better, the no-mixing gauge choice simplifies the form of the photon propagator so that it is more easily to study the UV divergence in radiative corrections and other physical phenomena. In the same way the fact that the minimum coupling of DKP scalar particles with Podolsky has just one vertex like QED, instead of 2 vertices when using KGF equations, decreases the number of Feynman diagrams to calculate radiative corrections. With the functional generator of GSDKP besides studying the Dyson-Schwinger covariant quantum equations and Ward-Takahashi identities from quantum gauge symmetry, we could establish the multiplicative renormalization procedure in the mass shell. An explicit calculation of the first radiative corrections (1-loop) associated with the photon propagator, meson propagator, vertex, photon-photon four point function utilizing the dimensional regularization method, where the gauge symmetry is manifest, will show two ways of evaluating the renormalization conditions for the pole and residue, due to DKP trilinear algebra, and we will also see that the DKP algebra ensures the functioning of the Ward-Takahashi (WT) identities in the first radiative corrections of the vertex and photon-photon four point function prohibiting UV divergences \cite{Andrenorm}.

Secondly, in constructing the partition function, it was necessary to study separately the matter, radiation and ghost sectors to later unify them. When we construct the scalar sector, we saw that DKP algebra came naturally in the Euclidean space and due to the global charge conservation the density matrix has a chemical potential, so the description of the interaction between the scalar particles and an external field is written in the grand canonical ensemble. On the other hand, in the vectorial sector due to the fact that there is no charge for the photons, the description of the interaction between the photons and external currents is written in the canonical ensemble. But as we are working with a thermal covariant language we do not have a way out, there are ghosts, they have charge and in the no-mixing gauge condition they are described by grasmanian and scalar fields due to the peculiar pseudo-diferential structure. So we unify all the structure in a variational principle of Gibbs entropy. Consequently, in writing the grand canonical potential, we have seen that physical degrees of freedom are closely related to energy equipartition theorem. There are 2 degrees of freedom for the scalars mesons, 8 degrees of freedom for the covariant Podolsky photons, 3 degrees of freedom for the ghosts and some interesting occurs, the ghost eliminate the degrees of freedom of the photons maintaining the physical degrees of freedom for Podolsky photons, which in this case are 5. With the partition function of GSDKP we could analyse the Schwinger-Dyson-Fradkin equations and Ward-Takahashi-Fradkin guage identities. In the same way, we could study the complete thermal quantum equations and gauge symetries in the Heisenberg description. We could then make explicit calculations of the first radiative thermal corrections and established the multiplicative renormalization procedure \cite{AndPime}.

With the previous results we can establish a set of rules that relates the quantum description and the thermal description of a field theory, in terms of a Wick rotation to imaginary time

\begin{eqnarray*}
&&\text{(Minkowski space)}\;(t,\vec{x})\rightarrow(i\tau,\vec{x})\;\text{( Euclides space)} \cr\cr
&&\Box\rightarrow-\triangle\cr\cr
&&\int d^{4}x\rightarrow i\int_{0}^{\beta}d\tau\int d^{3}\vec{x}\cr\cr
&&D_{\mu}=\partial_{\mu}-ieA_{\mu}\rightarrow D_{\mu }^{(e,\mu _{e})}=\partial _{\mu }-ie{\mathcal{A}}_{\mu}^{E}-\mu _{e}\delta _{\mu 0}.
\end{eqnarray*}
as seen in the link between eq. (\ref{qed1}) and eq. (\ref{qed2}) or eq. (\ref{funcgerador}) and eq. (\ref{parteq}) respectively. The concept that guarantees this link between the theories $(T = 0 | T \neq 0)$ are the similarity transformations.

\section{Acknowledgement}

A. A. Nogueira thanks (PNPD/Capes-UFABC) for support, L. Rabanal thanks CAPES for support and B. M. Pimentel thanks CNPq for partial support.

\end{document}